\documentclass[aps]{revtex4}

\usepackage{graphicx}
\usepackage{amsmath}
\usepackage{amsfonts}
\begin{document}
\begin{flushright}
INT-
\end{flushright}
\baselineskip=18pt

\title{\bf  Convergence properties of {\itshape ab initio} calculations of light nuclei in a harmonic oscillator basis}
\author{ S.~A.~Coon$^a$, M.~I.~Avetian$^a$, M.~K.~G.~Kruse$^a$,  U.~van Kolck$^{a,b}$, P.~Maris$^c$, J.~P.~Vary$^{c}$  \\
\it $^a$ Department of Physics, University of Arizona, Tucson, Arizona 85721\\
\it $^b$ Institut de Physique Nucl\'{e}aire, Universit\'{e} Paris-Sud, IN2P3/CNRS, F-91406 Orsay Cedex, France\\
\it $^c$ Department of Physics and Astronomy, Iowa State University, 
		Ames, IA 50011}

\maketitle
\vskip5mm

\noindent
{\bf Abstract}

We study recently proposed ultraviolet and infrared momentum regulators of the model spaces formed by  construction  of a variational trial wavefunction which uses a complete set of many-body basis states  based upon three-dimensional harmonic oscillator (HO) functions.  These model spaces are defined by  a truncation of the expansion characterized by a counting number ($\mathcal{N}$) and by  the intrinsic scale  ($\hbar\omega$) of the HO  basis; in short by the ordered pair ($\mathcal{N},\hbar\omega$).  In this study we  choose for $\mathcal{N}$ the truncation parameter $N_{max}$ related to the maximum number of oscillator quanta, above the minimum configuration, kept in the model space. The ultraviolet (uv) momentum cutoff of the  continuum is readily mapped onto  a defined uv cutoff in this  finite model space, but there are two proposed definitions of the infrared (ir) momentum cutoff inherent in a finite-dimensional HO basis.  One definition  is based upon the lowest momentum difference given by  $\hbar\omega$ itself and the other upon the infrared momentum which corresponds to the maximal
radial extent  used to encompass the many-body system in coordinate space.  Extending both the uv cutoff to infinity and the ir cutoff to zero  is prescribed for a converged calculation.  We calculate the ground state energy of light nuclei with ``bare" and ``soft" $NN$ interactions. By doing so,  we investigate the behaviors of the uv and ir regulators of model spaces used to describe  $^2$H, $^3$H, $^4$He and $^6$He with $NN$ potentials Idaho N$^3$LO and JISP16.  We establish practical procedures which utilize these regulators to obtain the extrapolated  result from sequences of calculations with model spaces characterized  by ($\mathcal{N},\hbar\omega$).

\maketitle

\section{Introduction}
\label{intro}

It has long been suggested that the three-dimensional (3d) harmonic oscillator (HO) provides a suitable expansion basis for a straightforward variational calculation of the properties of light nuclei.  In a traditional variational calculation, a trial wavefunction is selected having a form which aims to exploit all of the important features of the Hamiltonian under investigation, and its parameters are adjusted to minimize the energy of the few body system \cite{MCreview}.  It is appealing to generate a trial wavefunction in a completely systematic manner without regard for the details of the Hamiltonian under consideration other than the implementation of exact symmetries.  The goal, then, is to define a complete set of states for a few-body system and to construct and diagonalize the Hamiltonian matrix in a truncated basis of these states. The result of the diagonalization is an upper bound to the exact eigenvalue of the complete set.  With this method, in contrast to that of a pre-chosen trial wavefunction expected to capture the physics, a reliable estimate of the accuracy attained can be made with the variational upper bound \cite{Delves72} provided that the trial function is constructed using the terms of a systematic expansion set and convergence of the diagonalization result (such as a ground-state energy) is observed as the basis is increased.

The algebra appropriate to generating and using trial wavefunctions, based on 3d HO eigenfunctions, has been given by Moshinsky \cite{Moshinsky} and others \cite{others}.  The trial functions take the form of a finite linear expansion in a set of known functions 
 \[\Psi_T = \sum_{\nu}a_\nu^{({\mathcal{N}})}h_\nu  \]
 where $a_\nu^{({\mathcal{N}})}$ are the parameters to be varied and  $h_\nu$ are many-body states based on a summation over products of HO functions.   The advantage of a HO basis is that it is relatively straightforward to construct a complete set of few-body functions of appropriate angular momentum and symmetry; examples are given in Refs. \cite{others,JLS70}.
 The trial function must have a definite symmetry reflecting the  composition of the bound state: fermions or bosons.   This trial function $\Psi_T$ must be quadratically integrable and the expectation value of the Hamiltonian must be finite.   The expansion coefficients (known as generalized Fourier coefficients in the mathematical literature) 
depend on the upper limit (such as an $\mathcal{N}$ defined in terms of total oscillator quanta) and are obtained by minimizing the expectation value of the  Hamiltonian in this basis. 
Treating the coefficients $a_i^{({\mathcal{N}})}$ as variational parameters in the Rayleigh quotient  \cite{Kruse}, one performs the variation by diagonalizing the many-body Hamiltonian in this basis.
This is an eigenvalue problem so the minimum with respect to the vector of expansion coefficients always exists and one obtains a bound on the lowest eigenvalue.  The basis functions can also depend upon a  parameter (such as the harmonic oscillator energy $\hbar\omega$ which sets a scale)  that then becomes a  non-linear variational parameter additional to the linear expansion coefficients.    Such variational approaches were the standard  for  calculating properties of the trinucleon in the decade following the 1960's \cite {JLS70,Ciofi85} and have also been    applied to three and four body  alpha particle models of light nuclei and hypernuclei, see \cite{PC79,PC91,Portilho02}). 
No-core shell model (NCSM) calculations and no-core full configuration (NCFC) calculations with a ``bare" potential are  more recent examples of a variational calculation with a linear trial wave function.  Here the basis truncation parameter $\mathcal{N}$ and the HO energy parameter $\hbar\omega$ are variational parameters \cite{NavCau04, Maris09,NQSB}.

 In such a calculation one would like not only to obtain rapid convergence of the eigenvalue and wavefunction but one would like this convergence to be to the exact solution.   The functional analysis theorems needed for the discussion of the convergence properties of a linear trial function are displayed in Appendix A of the article ``Variational Techniques in the Nuclear Three-Body Problem" \cite{Delves72}.   We quote from Section 2.3.5  of the article: ``It is shown there [Appendix A] that, provided the set of expansion functions is suitably complete [ i.e. complete in the energy norm], one will eventually obtain convergence [of the lowest approximate eigenvalue] to the exact value [by increasing the basis]. Moreover, if the set is constructed systematically, then in general one can expect the convergence to be smooth; indeed, we can often predict the rate at which the convergence will occur as we shall discuss... In these circumstances, the numerical convergence of the upper bound can provide a useful estimate of the accuracy of the calculation, and one which is in practice very much more realistic than that derived from the direct lower bound calculation."   With a linear trial function, the expectation value W($\mathcal{N}$) of any bounded operator W will converge provided that the energy converges; and one may estimate the accuracy obtained by watching the numerical convergence of W($\mathcal{N}$) with increasing $\mathcal{N}$  \cite{bounded}.  Such examinations (albeit for rather small basis size compared to those used in this study) are displayed in Ref. \cite{CP87}.

The {\it rate} of convergence and the number of terms needed for this eventual asymptotic rate to ``start to behave"  is of great practical importance for extrapolation \cite{Schwartz}. This question is discussed at great length by Delves \cite{Delves72} with general theorems and numerical examples for smooth (e.g. attractive Gaussian which is finite everywhere) and non-smooth (e.g. attractive Yukawa which has a singularity at the origin)  local two-body potentials and a variety of trial functions. As an example, Delves derives for  the harmonic oscillator basis a convergence rate  according to the inverse squared power of  $\mathcal{N}$ for ``nonsmooth" potentials such as a Yukawa; a convergence rate expected to be independent of the number of particles.  He then demonstrates that the  binding energies in the truncated expansions  of Ref.  \cite{JLS70}   do 
follow this power law for both the deuteron of the Reid soft-core potential \cite{Reid} (the archetype of a sum of Yukawas with strong high-momentum components) and the deuteron and triton of the (separable) Yamaguchi potential \cite{Yamaguchi}.   This is very slow convergence compared to other  sets of expansion functions popular in atomic and molecular physics and physical chemistry;  see Table V of Ref. \cite{Delves72}. Slow convergence hinders progress either because the amount of computation needed to reach a desired accuracy is prohibitive, or because too many arithmetic operations cause excessive round off error \cite{CP87,Portilho02,MP}.  Indeed, the slow convergence of {\it systematic} expansions was likely a contributing factor to the replacement of variational methods by finite difference methods (based upon the Faddeev decomposition) in the 1970s to treat the three-nucleon bound state problem.

 In a parallel application of functional analysis to a variational calculation by expansion in a basis, specific theorems about the asymptotic rate of convergence for the three-body bound state were developed by Schneider for a general basis \cite{Sch72}.  The conclusion was that ``In any particular problem the precise rate will depend on the exact form of the Hamiltonian and the operators [which determine the set of basis states] chosen".  The practical application of that paper was to the hyperspherical harmonics (HH) basis using simple schematic two-body potentials. The asymptotic rate of convergence of the three-body binding energy was suggested to converge as the inverse fourth power of the maximal grand angular quantum number $K$ for a Yukawa potential and exponentially fast in $K$   for a Gaussian potential   \cite{Sch72}.   These theorems were illustrated by explicit HH calculations of the 1970's \cite{Fabre}. The general expectations of these theorems continue to backstop  extrapolations in contemporary few-body calculations with modern potentials using this HH method \cite{Barnea1,Barnea2,Bacca}.  As suggested in Ref. \cite{Sch72}, the rate of convergence does not depend on the number of particles in the bound state. (This analysis was for 3- and 4-body systems which have very high first breakup thresholds-the rate of convergence is, however, expected to depend on the first breakup threshold of heavier nuclei).  Indeed, contemporary HH analyses of the four-nucleon bound state bear out this general expectation, although additional criteria for selecting a reduced basis have to be specified, and the authors of  \cite{Viviani05} demonstrate  that the inverse power law in $K$ can be higher than four for contemporary ``nonsmooth" two-  and three-body potentials.

 We are unaware of an application of the theorems proved by Schneider \cite{Sch72} to the HO basis.  However, a very up-to-date discussion of the full configuration-interaction (CI) method in a HO basis  does analyze convergence and gives practical convergence estimates for many-electron systems trapped in a harmonic oscillator (a typical model for a quantum dot) \cite{Kvaal}.  A corresponding investigation of light nuclei with another CI method, the NCFC approach \cite{Maris09}, provides consistent and tested uncertainty estimates for ground state energies. 
The CI method consists of approximating eigenvalues of the many-body Hamiltonian with those obtained by projecting the problem onto a finite dimensional subspace of the full Hilbert space  and diagonalizing the Hamiltonian in this model space \cite{Szabo}. Mathematically, this is analogous to a Ritz-Galerkin method on the model space spanned by the basis functions and the analysis of the energy error is equivalent to analysis of the corresponding Raleigh-Ritz calculation sketched earlier.  The projection can either take the form of an $\mathcal{N}$ defined in terms of total oscillator quanta (called ``total-energy-cut space" ) or in the single-particle quantum numbers (called ``single-particle-cut space").  The total-energy-cut space is used in this study (see Section II) and the latter single-particle-cut space lends itself more readily to approximating a full CI calculation by a coupled cluster approach.  The CI approach  becomes, in principle, exact as $\mathcal{N}\rightarrow \infty$ with either choice of $\mathcal{N}$.   For this reason the CI approach with HO basis functions is sometimes called called ``exact diagonalization".  A succinct statement of the equivalence of large-scale diagonalization and the Rayleigh-Ritz variational method can be found in the Introduction of Ref. \cite{Kvaal2}  and the full discussion in the monograph Ref. \cite{Gould}.

 As in the nuclear examples \cite{Delves72,Sch72}, the asymptotic convergence of the lowest eigenfunction of a quantum dot in the ``total-energy-cut" model  space  is slow and it is slow for the analogous  reason:  the singularity of the Coulomb interaction at those points where two or more interparticle distances are zero (the Kato cusp condition on the many-body wavefunction \cite{Kato}).  The convergence rate is dominated by the singularities in the analytic structure of the solution \cite{Morgan}: it takes many HO eigenfunctions to approximate the singularities of the many-body wavefunction due to the two-body interaction.  The asymptotic convergence of the nuclear structure problem is not changed by including Jastrow type two-body correlation functions in the trial wavefunction \cite{CP87}.  The onset of asymptotic convergence occurs, however, at a much smaller value of an $\mathcal{N}$ than for the case without correlation functions and the convergence is to the same final value of the lowest variational energy \cite{Ciofi85,CP87,Portilho02}.  Therefore much less computational resources are required to get the answer.

With the HO basis in the nuclear structure problem,  convergence has been discussed, in practice,  with an emphasis on obtaining those parameters which appear linearly in the trial function (i.e. convergence with $\mathcal{N}$).  Sometimes  for each $\mathcal{N}$ the  non-linear parameter $\hbar\omega$ is varied to obtain the minimal energy  \cite{CP87} and then the convergence with $\mathcal{N}$ is examined.   Sometimes $\hbar\omega$ is simply fixed at a value which gives the fastest convergence in $\mathcal{N}$ \cite{JLS70}.
 More recently, in the context of no-core shell model (NCSM) calculations and no-core full configuration (NCFC) calculations with smooth `bare' potentials, one sees figures or tables in which one of the variational variables of ($\mathcal{N},\hbar\omega$)  is held fixed and the variational energy displayed with respect to the other. This practice is helpful for the following reason.  Optimum values for the parameters that enter linearly can be obtained by solving a matrix eigenvalue problem. But the optimum value of the nonlinear parameter must in principle be obtained by, for example, numerical minimization which could be difficult as the algorithm could easily miss the global minimum and get trapped in a local minima.
The plots one sees in the nuclear physics  literature show that 1) for small bases a change in the non-linear parameter $\hbar\omega$ can make a dramatic change in the variational estimate of the ground state energy and 2) the dependence on the nonlinear parameter decreases as the basis size increases.  These observations seem to vitiate the need for an extensive numerical minimization by varying $\hbar\omega$ \cite{caveat}.  These observations have inspired definite (and differing) prescriptions  for convergence and extrapolation.
 It is the purpose of this study to suggest that effective field theory (EFT) concepts of ultraviolet (uv) and infrared (ir) cutoffs provide an alternative useful way to think about convergence and a physically motivated prescription for extrapolation of (necessarily truncated) results in the model space (elucidated in Section 2) of the trial wavefunction to the full Hilbert space.  

The paper is organized as follows. In section 2 we briefly describe  expansion schemes in HO functions. This expansion technique still retains the variational character described above.  We employ  realistic smooth nucleon-nucleon potentials (JISP16 \cite{Shirokov07} and Idaho N$^3$LO \cite{IdahoN3LO}) which have also been used by other authors without  renormalization  for $A\leq 6$ (Ref. \cite{Maris09} and Ref. \cite{NavCau04} respectively).  None of the discussion in section 2 is new, but it paves the way for section 3 in which we suggest a convergence analysis based upon the uv and ir cutoffs introduced in Ref. \cite{EFTNCSM} in the context of an EFT framework.   Section 4 is devoted to tests and examples of this new convergence scheme and section 5 contains a summary and outlook.

\section{Expansion in a finite basis of harmonic oscillator functions}
\label{sec:2}

Here we indicate the workings of the finite HO basis calculations performed and refer the reader to a very useful  review article \cite{NQSB} on the no-core shell model (NCSM) for further details and references to the literature.   In these no-core approaches, all the nucleons are considered active, so there is no inert core as in standard shell model calculations; hence the ``no-core" in the name.  $NN$ potentials with strong short-range repulsions and the concomitant high-momentum components do not lend themselves well to a HO basis expansion, as was well appreciated fifty years ago \cite{JLS70}.   A ``renormalization" of the Hamiltonian is often made by constructing an effective interaction (dependent upon the basis cutoff  $\mathcal{N}$ and upon $\hbar\omega$) by means of a unitary transformation due to Lee and Suzuki  \cite{NQSB}.  This  procedure generates effective many-body interactions that are often neglected \cite{NCSMC12}.  This neglect destroys the variational nature of a NCSM calculation.  We instead choose ``soft" potentials (JISP16 \cite{Shirokov07} and Idaho N$^3$LO \cite{IdahoN3LO})  which have also been used by other authors without  renormalization  for $A\leq 6$ (Ref. \cite{Maris09} and Ref. \cite{NavCau04} respectively), so that we can study convergence and extrapolation issues directly within a variational framework. NCSM calculations with these potentials are variational with the HO  energy parameter $\hbar\omega$ and the basis truncation parameter $\mathcal{N}$ as variational parameters \cite{NQSB}. Nomenclature has diverged somewhat since the advent of these smooth but still realistic potentials into a framework (NCSM) which originally included renormalization of the $NN$ potential.  Sometimes one reads about ``NCSM calculations with unmodified or `bare' potentials" \cite{NavCau04,NQSB}, or ``the no-core full configuration (NCFC) method"     \cite{Maris09}, or simply ``we use the basis of the no-core shell model (NCSM)" \cite{Jurgenson}.  All these phrases refer to retaining the original interaction (without renormalization) within the model space.   Nor do we renormalize the interaction in our study. 

To study these convergence issues we mostly employ the Idaho N$^3$LO $NN$ potential which is inspired by chiral perturbation theory and fits the two body data quite well \cite{IdahoN3LO}.
 It is  composed of contact terms and irreducible pion-exchange expressions multiplied by a regulator function designed to smoothly cut off high-momentum components in accordance with the low-momentum expansion idea of chiral perturbation theory.   The version we use has the high-momentum cutoff of the regulator set at 500 MeV/c.  The Idaho N$^3$LO potential is a rather soft one, with heavily reduced high-momentum components as compared to earlier realistic $NN$ potentials with a strongly repulsive core.  Alternatively, in coordinate space, the Yukawa singularity at the origin is regulated away so that this potential would be considered ``smooth" by Delves and 
Schneider and the convergence in $\mathcal{N}$ would be expected to be exponential \cite{Delves72, Sch72}.  Even without the construction of an effective interaction, convergence with the Idaho N$^3$LO $NN$ potential is exponential, as numerous studies have shown \cite{NavCau04,Jurgenson}.   Nevertheless, it has been useful to simplify and reduce the high-momentum components of this and other phenomenological potentials further by means of the similarity renormalization group evolution \cite{Jurgenson}.   Such a softening transformation is imperative for heaver nuclei ($A > 6$) and/or if three-nucleon forces are included in the Hamiltonian \cite{Bogner,Jurgenson,NAVQ}.


The second $NN$ interaction we employ is JISP16 \cite{Shirokov07}, a nonlocal separable potential whose form factors are HO wavefunctions.  It is constructed by means of the $J$-matrix version of inverse scattering theory. The matrix of the $NN$ potential in the oscillator basis is obtained for each partial wave independently, so the $NN$ interaction is a set of potential matrices for different partial waves \cite{ JISP16_web}.  These matrices reproduce
the experimental $NN$ scattering data and properties of the deuteron to high precision.  Once the inherent ambiguity of this method is eliminated by a plausible phenomenological ansatz, the scattering wavefunctions are very close to the ones provided by meson exchange
 ``second-generation" $NN$ potentials \cite{Shirokov04}.   As for the name of this potential, JISP refers to $J$-matrix Inverse Scattering Potential and version ``16" has had phase-equivalent unitary transformations applied to selected partial waves so that the resulting interaction continues to describe two-body data well.  Selected partial waves are tuned to provide good descriptions of $^3$H binding, the low-lying spectra of $^6$Li and the binding energy of $^{16}$O  \cite{Shirokov07}.  The virtue of this potential is that it is also  ``soft".  Although nonlocal and not really fitting into Delves classification, it is not surprising that variational calculations  with this $NN$ interaction also converge exponentially with $\mathcal{N}$  \cite{Maris09} since the HO form factors of this separable potential are gaussians multiplied by polynomials in the radial coordinate.   (It is noteworthy   that JISP16 in the HH basis also converges exponentially in  $K$ \cite{betadecay}, as would be expected by Schneider \cite{Sch72}).

 We use a HO basis that allows preservation of translational invariance of the nuclear self-bound system.  Translational invariance is automatic if the radial HO wavefunction depends on relative, or Jacobi, coordinates as was done in Refs. \cite{JLS70,Ciofi85,PC79,PC91,Portilho02}. Antisymmetrization (or symmetrization for the alpha particle models of \cite{Ciofi85,PC79,PC91,Portilho02}) of the basis is necessary and described in Refs. \cite{NQSB} and  \cite{NKB}.  Antisymmetrization in a Jacobi basis becomes analytically and computationally forbidding as the number of nucleons increases beyond  four or five.  For this reason  these calculations are alternatively made with antisymmetrized wavefunctions constructed as Slater determinants of single-nucleon wavefunctions depending on single-nucleon coordinates. This choice loses translational invariance since, in effect, one has defined a point in space from which all single-particle coordinates are defined. Translational invariance is  restored by using the ``Lawson method" \cite{Lawson} to be described shortly.   The gain of this choice is that one  can use   technology developed and/or adapted for NCSM, such as the  parallel-processor codes ``Many-Fermion Dynamics --- nuclear" (MFDn)
\cite{Vary92_MFDn} 
and the No-Core Shell Model Slater Determinant Code \cite{NCSMSD}.  These   codes set up the many-body basis space,
evaluate the many-body Hamiltonian matrix, 
obtain the low-lying eigenvalues and eigenvectors using the
Lanczos algorithm, and evaluate a suite of expectation values using the eigenvectors.

The Slater determinant basis is often defined in the  ``$m$-scheme" where each HO single-particle state has its orbital and spin
angular momenta coupled to good total angular momentum, $j$, and its magnetic
projection, $m$.  
The many-body basis states for a given total number of nucleons $A$ are Slater determinants in this HO basis and
are limited by the imposed symmetries --- parity, charge 
and total angular momentum projection ($M$), as well as by $\mathcal{N}$.   
In the natural parity cases for even nuclei $M=0$,
enables the simultaneous calculation of the entire low-lying  spectrum for that
parity and the chosen $\mathcal{N}$.

The use of this specially constructed Slater determinant basis results in eigenstates of a translationally invariant Hamiltonian (supplemented by a suitable constraint term) that factorize as products of a wavefunction depending on relative coordinates and a wavefunction depending on the CM coordinates.  This is true for a particular truncation of the basis: a maximum of the sum of all HO excitations, i.e. $\sum_{i=1}^A (2n_i + l_i) \leq N_{totmax}$, where $n_i,l_i$ are the HO quantum numbers corresponding to the harmonic oscillators associated with the single-nucleon coordinates and $N_{totmax}$ is an example of the generic $\mathcal{N}$ of the Introduction.   Note that this truncation is on the level of total energy quanta (``total-energy-cut space"), which is different  from the CI calculations  used in atomic and molecular problems, which are often truncated on the single-particle level (``single-particle-cut space"). 

The precise method of achieving the factorization of the CM and intrinsic
components of the many-body wavefunction follows a standard approach, sometimes
referred to as the ``Lawson method" \cite{Lawson}.  
In this method, one selects the many-body basis space in the manner described above 
with $\mathcal{N}=N_{totmax}$
and adds
a Lagrange multiplier term to the many-body Hamiltonian 
$\beta(H_{CM} - \frac{3}{2} \hbar\omega)$
where $H_{CM}$ is the HO Hamiltonian for the CM motion.  With $\beta$
chosen positive (10 is a typical value), one separates the states of lowest CM motion 
($0S_{\frac{1}{2}}$)
from the states with excited CM motion by a scale of order $\beta \hbar\omega$.  The 
resulting low-lying states have wavefunctions that then have  the desired factorized form. 
We checked, for the two cases $A=3$ and $A=4$, that the codes $\it manyeff$  \cite{NKB} which use Jacobi coordinates and  No-Core Shell Model Slater Determinant Code  \cite{NCSMSD} based upon single-nucleon coordinates
gave the same eigenvalues for the same values of  $\mathcal{N}=N_{totmax}$ and $\hbar\omega$, indicating that the Lawson method is satisfactory for the calculations in   single-particle coordinates.  Some details of this check will be given in  Section IV which gives results.

	Now we continue the discussion of  the (total-energy-cut space) truncation parameter $\mathcal{N}$ of the HO basis expansion of the many-body system.  Usually, instead of truncating the sum of all HO excitations 
	$\mathcal{N}= N_{totmax} \geq \sum_{i=1}^A (2n_i + l_i)$, 
	one introduces  the truncation parameter  $N_{max}$.
$N_{max}$ is the maximum number of oscillator quanta
shared by all nucleons above the lowest HO configuration allowed by the Pauli-exclusion principle for the chosen nucleus.   We label the HO shells by energy quanta $N=(2n+l)$, where $n=0,1,2, \ldots$ and $l=0,1,2, \ldots$.  
Thus, for example, in $^6$He a truncation at $N_{max}=4$ would allow one neutron to occupy the $N=5$ HO shell, the other ``valence" neutron would remain in the $N=1$ shell, and the remaining 4 nucleons remain in the (filled)  first shell labeled by $N=0$.  Alternatively, the two valence neutrons could occupy the $N=3$ shell and the remaining 4 nucleons stay in the $N=0$ shell. In both cases (and for all other combinatorics) in  $^6$He $N_{max} = N_{totmax}-2$.  Similarly for other 
 $p$-shell nuclei one can work out that  $N_{totmax}$ and $N_{max}$ differ, e.g. 
 for $^{12}C$,   $N_{max} = N_{totmax}-8$, etc. 
  However, for the s-shell nuclei $^2$H, $^3$H,$^3$He, and $^4$He $N_{max} = N_{totmax}$.   

Later on we will want to identify  parameters of the model space (with the dimensions of  momenta) which refer, not to the many-body system, but to the properties of the HO single-particle states.  If the highest HO single-particle (SP) state of the lowest HO
configuration  allowed by the Pauli-exclusion principle has $N_0$ HO quanta, then $N_{max}+N_0=N$.   Since
$N_{max}$ is the maximum of the {\em total} HO quanta above the
minimal HO configuration, we can have at most one nucleon in such a
highest HO SP state with $N$ quanta.  Note that $N_{max}$ characterizes the many-body basis space, whereas $N$ is a label of the corresponding highest single-particle orbital.  To find the value of the single-particle  label $N$, we need to determine the highest occupied SP state in a given $N_{max}$ truncation.  One gives {\it all} the available $N_{max}$ quanta to  a single nucleon.  Consider again a $^6$He basis truncated at $N_{max}=4$;  both valence neutrons occupy the $0p$ $(N_0=1)$ shell in the lowest energy many-body configuration.  Assigning a single neutron  the entire $N_{max}=4$ quanta means that, as before, the highest occupied SP state is in the $N=5$ shell.
 On the other hand, the highest occupied orbital of the closed s-shell nucleus $^4$He has $N_0=0$  so that $N=N_{max}$.

\section{convergence in uv and ir variables}
\label{sec}

We begin by thinking of the finite single-particle basis space defined by $N$
and $\hbar\omega$ as a model space characterized by two momenta associated with the basis functions themselves.
In the HO basis, we define $\Lambda=\sqrt{m_N(N+3/2)\hbar\omega}$ as the 
momentum (in units of MeV/c) associated with the energy of the highest HO level.  The nucleon mass is $m_N=938.92$ MeV. To arrive at this definition one applies the virial theorem to this highest HO level  to establish  kinetic energy as one half the total energy (i.e., $(N+3/2)\hbar\omega\:$)  and solves the non-relativistic dispersion relation for $\Lambda$.  This sets one of two cutoffs for the model space of a calculation.  Energy, momentum and length scales are related, according to Heisenberg's uncertainty principle. The higher the energy or momentum scale we may reach, the lower the length scale we may probe. Thus, the usual definition of an ultraviolet cutoff $\Lambda$ in the continuum has been extended to discrete HO states. It is then quite natural to interpret the behavior of the variational energy of the system with addition of more basis states as the behavior   of this observable with the variation of the ultraviolet cutoff $\Lambda$.  Above a certain value of $\Lambda$ one expects this running of the observable with $\Lambda$ to ``start to behave" so that this behavior can be used to extrapolate to the exact answer.    However, the model space has another scale which  motivates a second cutoff; the energy scale of $\hbar\omega$ itself.   Because the energy levels of a particle in a HO potential are quantized in units of $\hbar\omega$,  the minimum allowed momentum difference between single-particle orbitals  is 
$\lambda=\sqrt{m_N\hbar\omega}$ and that has been taken to be an infrared cutoff \cite{EFTNCSM}.  That is, there is a low-momentum cutoff $\lambda=\hbar/b$ corresponding to the minimal accessible non-zero momentum (here $b=\sqrt{\frac{\hbar}{m_N\omega}}$ plays the role of a characteristic  length of the HO potential and basis functions).  Note however that there is {\it no} external confining HO potential in place.  Instead the only  $\hbar\omega$ dependence is due to the scale parameter of the underlying HO  basis.

 The energies of a many-body system in the truncated model space will differ from those calculated as the basis size increases without limit ($N\rightarrow\infty$).  This is because the system is in effect confined within a finite (coordinate space) volume characterized by the finite value of $b$ intrinsic to the HO basis. The ``walls"  of the volume confining the interacting system spread apart  and the volume increases to the infinite limit  as $\lambda\rightarrow 0$ and $b\rightarrow\infty$. Thus it is as necessary to extrapolate the low momentum  results obtained with a truncated basis with a given $b$ or $\hbar\omega$ as it is to ensure that the ultraviolet cutoff is high  enough for a converged result.  These energy level shifts in a large enclosure have long been studied \cite{FukudaNewton}; most recently with the explicit EFT calculation of a triton in a cubic box allowing the edge lengths to become large (and the associated  ir cutoff due to momentum quantization in the box going towards zero) \cite{Kreuzer}.  There it was shown that as long as the infrared cutoff was small compared to the ultraviolet momentum cutoff appearing in the ``pionless" EFT, the ultraviolet behavior of the triton amplitudes was unaffected by the finite volume.  More importantly, from our point of view of desiring extrapolation guidance, this result means that calculations in a finite volume can confidently be applied to the infinite volume (or complete model space) limit. Similar conclusions can be drawn from the ongoing studies of systems of two and three nucleons trapped in a HO potential with interactions from pionless EFT and this definition ($\lambda=\sqrt{m_N\hbar\omega}$) of the infrared cutoff \cite{EFTtrapped}.

Other studies define the ir cutoff as  the infrared momentum which corresponds to the maximal
radial extent  needed to encompass the many-body system we are attempting to describe by the finite basis space (or model space).  These studies find it natural to define the ir cutoff by $\lambda_{sc}=\sqrt{(m_N\hbar\omega)/(N+3/2)}$ \cite {Jurgenson,Papenbrock}.  Note that $\lambda_{sc}$ is the inverse of the root-mean-square (rms) radius of the highest single-particle state in the basis; $\langle r^2\rangle^{1/2}=b
\sqrt{N+3/2} $.   We distinguish the two definitions by  denoting the first (historically) definition by $\lambda$ and the second definition by $\lambda_{sc}$ because of its  scaling properties demonstrated in the next Section. This latter ir variable $\lambda_{sc}$ clearly goes to zero either i) as $\hbar\omega$ goes to zero at fixed $N$ or ii) as $N$ becomes large for fixed $\hbar\omega$.  It is the second limit which corresponds to the convergence theorems of the Introduction.  In this latter limit 
$\lambda_{sc}\rightarrow 0$ while $\lambda$ does not.

 The extension of the continuum ultraviolet cutoff to the discrete (and truncated) HO basis with the definition  $\Lambda=\sqrt{m_N(N+3/2)\hbar\omega}$ seems unexceptional.  An equally plausible alternative uv cutoff differs from $\Lambda$ by only a scale change \cite{Kalliofootnote}, in striking contrast to the alternate definitions of the ir cutoff which have different functional forms.  It is a goal of this work to determine the usefulness  of the two rival definitions-$\lambda$ and $\lambda_{sc}$-of the infrared regulator (let us call it $\lambda_{ir}$) of the model space with parameters  ($N$, $\hbar\omega$).   From the beginning, it is clear that increasing $\Lambda$   by increasing $\hbar\omega$ in a fixed-$N$ model space is not sufficient; doing so   increases both of the putative infrared cutoffs as well because  $\Lambda = \lambda\sqrt{N+3/2} = \lambda_{sc}(N+3/2)$ and one continues to effectively  calculate in an effective confining volume which is getting smaller rather than larger.  This confining volume is certainly removed by letting  $N\rightarrow\infty$, at fixed $\hbar\omega$, because HO functions form a basis of the complete space.  But increasing $N$ without limit  is computationally  prohibitive, and furthermore will not shed much light on the question of whether in  practical calculations one must take the ir cutoff to zero by taking $\hbar\omega\rightarrow 0$  ($\lambda_{ir}\equiv\lambda$ definition) or whether it is sufficient to allow $\hbar\omega$ be some larger value, perhaps near that used in  traditional shell-model calculations, and let an increasing $N$ take $\lambda_{ir}$ to small values, as it does with the definition $\lambda_{ir}\equiv \lambda_{sc}$.
 
\newpage 
\begin{figure}[htpb]
\includegraphics[width=16cm,height=12cm,clip]{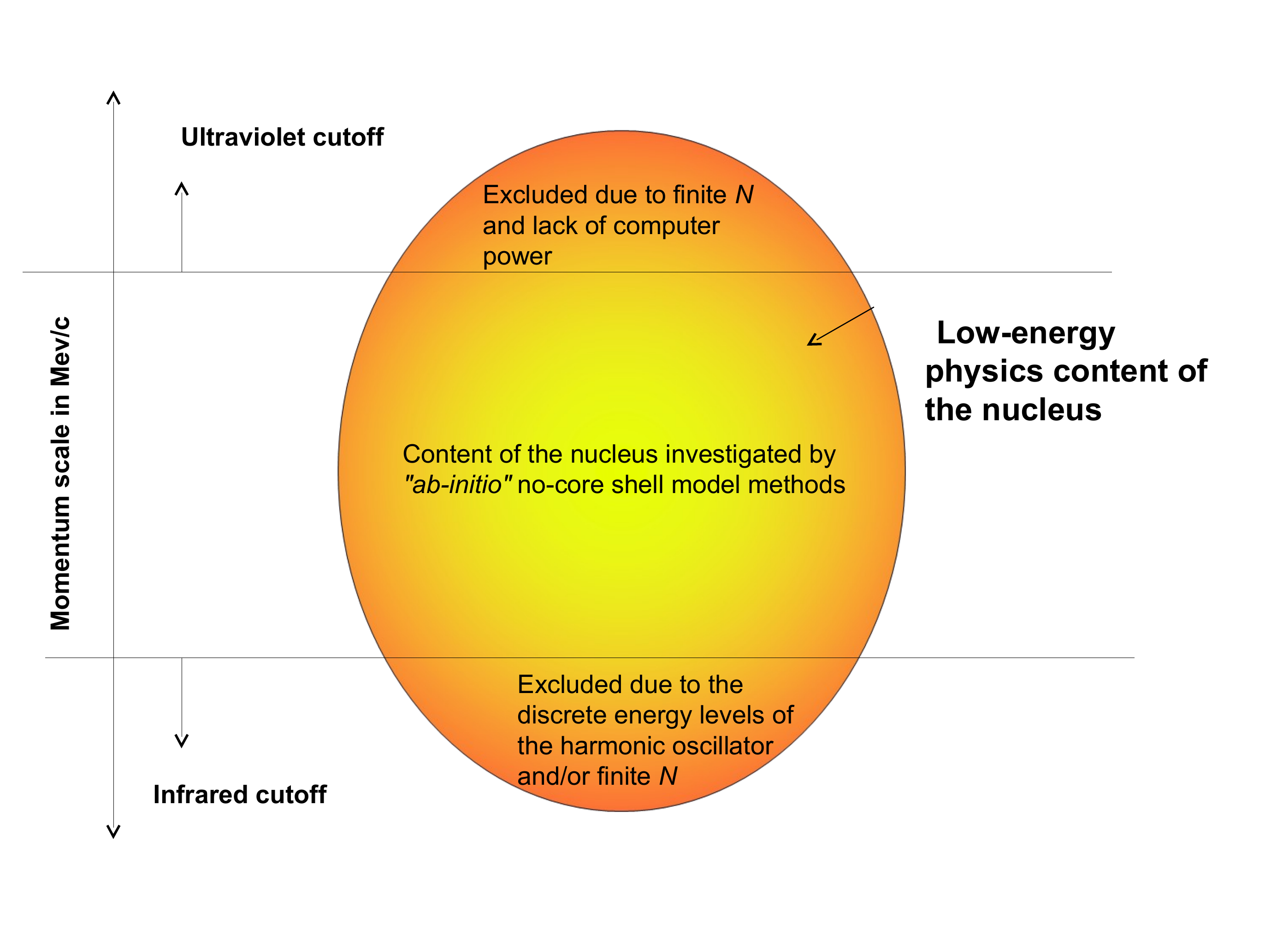}
\caption{(Color online)  Schematic view of a finite model space (limited by the basis truncation parameter $N$ as described in the text), in which the uv and ir momentum cutoffs are arbitrary.  To reach the full many-body Hilbert space, symbolized by the complete oval,  one needs to let the uv cutoff $\rightarrow \infty$ and the ir cutoff $\rightarrow  0$.}

\label{fig:1}
\end{figure}

We are interested in the limit of large $\Lambda$ and small $\lambda_{ir}$ (see Figure 1).   If one can establish that the cutoff dependences of the model space decrease with increasing $\Lambda$ and decreasing $\lambda_{ir}$ then one can i) remove the influence of the ir cutoff by extrapolating to the infrared limit for selected uv-cutoff values chosen to be above the uv nature of the potential and ii) if needed,  extrapolate to the uv limit for selected ir cutoff values chosen by the size of the system modeled.  We will show that such a program is possible.

\section{Results and Discussion}
\label{sec:4}

\begin{figure}[htpb]
\includegraphics[width=16cm,height=12cm,clip]{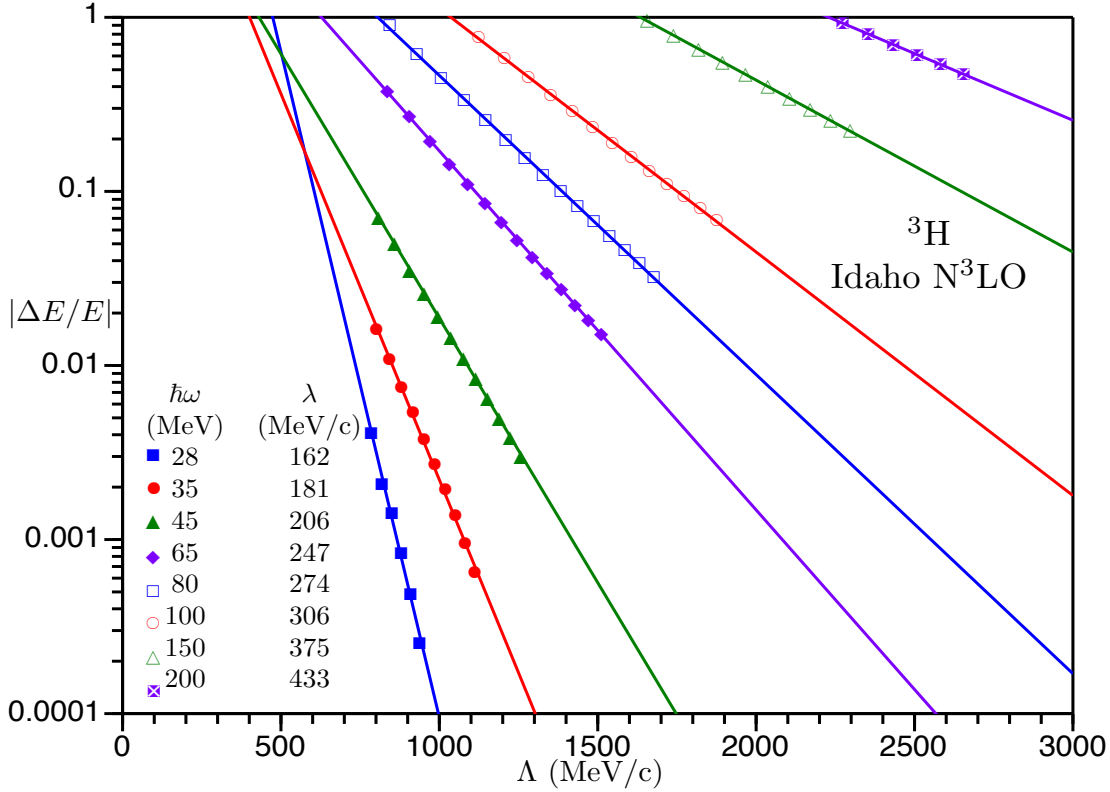}
\caption{(Color online)  Dependence of the ground-state energy of $^3$H (compared to a converged value-see text) upon the uv momentum cutoff $\Lambda$ for different fixed $\lambda$.  The curves are fit to the calculated points.}

\label{fig:2}
\end{figure}

We first display in a series of figures the running of the ground-state eigenvalue of a single nucleus, $^3$H, on the truncated HO basis by holding one cutoff of ($\Lambda,\lambda_{ir}$) fixed and letting the other vary.   Then we show that the trends noted hold for other light nuclei within the range of our computer resources.    Finally we discuss extrapolation procedures. 

These $^3$H calculations were made for $N\leq 36$ and  values of $\hbar\omega$ as appropriate for the chosen cutoff value.  For $N\geq 16$, we used the code $\it manyeff$  \cite{NKB}  which uses  Jacobi coordinates and the No-Core Shell Model Slater Determinant Code \cite{NCSMSD}  which use single-particle coordinates for smaller $N$.   We checked that the codes gave the same eigenvalues for overlapping values of $N$, indicating that the Lawson method  satisfactorily restores translational invariance to ground-state energy calculations in   single-particle coordinates.  For example, the ground state energy of $^3$H with the  Idaho N$^3$LO $NN$ potential at $N_{max} =16$ and $\hbar\omega = 49.2968
$ MeV is $( -7.3378,-7.3385)$ MeV for the (Jacobi, single-particle) basis choice.

In Figure 2 and the following figures,  $\vert\Delta E/E\vert $ is defined as $\vert (E(\Lambda,\lambda_{ir}) - E)/E\vert$ where $E$ reflects a consensus ground-state energy from benchmark calculations with this $NN$ potential,  this nucleus, and different few-body methods. The accepted value for the ground state of  $^3$H with this potential is $ -7.855$ MeV from a 34 channel Faddeev calculation \cite{IdahoN3LO}, $ -7.854$ MeV from a hyperspherical harmonics expansion \cite{Kievsky08}, and $-7.85(1)$ from a NCSM calculation \cite{NavCau04}.
 All $\vert\Delta E/E\vert $, starting with Figure 2, will follow some trajectory (trajectory's shape not predicted). 
For the choice of Figure 2, $\lambda_{ir}\equiv \lambda=\sqrt{m_N \hbar\omega}$, \ $\vert\Delta E/E\vert \rightarrow 0$ as $\Lambda$ increases.  Fixed $\hbar\omega$ implies $N$   $\it{alone}$ increases to drive $\Lambda\rightarrow\infty$,  $\lambda_{sc}\rightarrow 0$ simultaneously.   The linear fit on a semi-log plot is extracted from the data.  This fit implies $\vert\Delta E/E\vert  \sim B\exp(-\Lambda/\Lambda_{ref}(\lambda))$,  where $B$ is approximately constant and $c \Lambda_{ref}(\lambda)\sim 30 \hbar\omega$ for $\hbar\omega > 45$ MeV. Note this is $\Lambda_{ref}(\lambda)$ and not  $\Lambda_{ref}(\lambda_{sc})$, i.e. with fixed $\lambda$,  $\Lambda_{ref}$ is a constant.
 On the other hand, 
 for fixed $\Lambda$, a smaller $\lambda$ implies a smaller 
 $\vert\Delta E/E\vert$ since more of the infrared region is included in the calculation.\\
 

\begin{figure}[htpb!]
\includegraphics[width=16cm,height=12cm,clip]{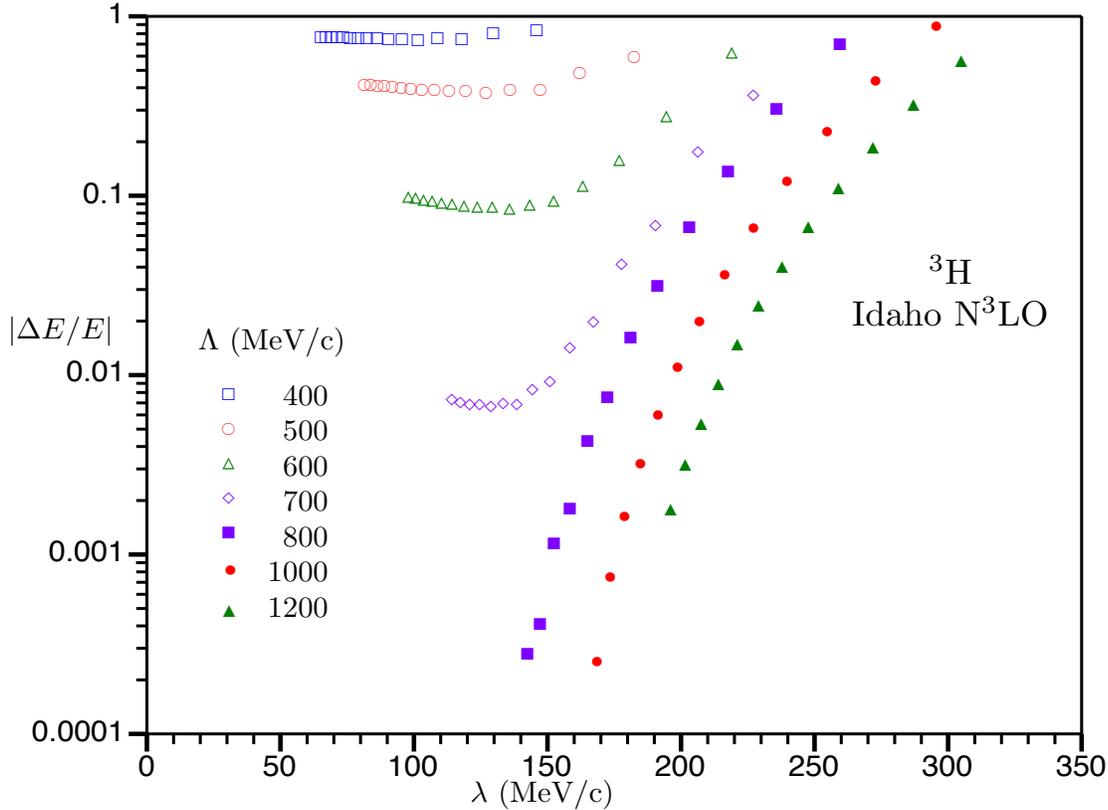}
\caption{(Color online)  Dependence of the ground-state energy of $^3$H (compared to a converged value-see text) upon the ir momentum cutoff $\lambda$ for fixed $\Lambda$.  }
\label{fig:3}
\end{figure}

In Figure 3 we hold fixed the uv cutoff of ($\Lambda,\lambda_{ir}$) to display the running of  $\vert\Delta E/E\vert$ upon the suggested ir cutoff $\lambda$.   For fixed $\lambda$, a larger $\Lambda$ implies a smaller  $\vert\Delta E/E\vert$ since more of the uv region is included in the calculation.  
But we immediately see a qualitative change in the curves between the transition $\Lambda=700$ MeV and $\Lambda=800$ MeV; for smaller $\Lambda$, $\vert\Delta E/E\vert$ does not go to zero as the ir cutoff is lowered and more of the infrared region is included in the calculation.  This behavior suggests that $\vert\Delta E/E\vert$ does not go to zero unless $\Lambda\geq\Lambda^{NN}$, where
 $\Lambda^{NN}$ is some uv regulator scale of the $NN$ interaction itself.  From this figure one estimates  $\Lambda^{NN}\sim$ 800 MeV/c for the Idaho N$^3$LO interaction. 
 
 Yet the description of this interaction in the literature says that the version we use has the high-momentum cutoff of the regulator set at $\Lambda_{N3LO} = 500$ MeV/c \cite{IdahoN3LO}.  This does not mean that the interaction has a sharp cutoff at exactly  500 MeV/c, since the terms in the Idaho N$^3$LO interaction are actually regulated by an exponentially suppressed term of the form 
 
 \[  \exp  \left [ - \left( \frac{p}{\Lambda_{N3LO}}\right) ^{2n} - \left( \frac{p'}{\Lambda_{N3LO}}\right) ^{2n}\right] .    \]
In this expression, $p$ and $p'$ denote the magnitude of the  initial and final nucleon momenta of this non-local potential in the center-of-mass frame and $n\geq 2$.  Because the cutoff is not sharp, it should not be surprising that one has not exhausted the uv physics of this interaction for values of single-particle $\Lambda$ somewhat greater than 500  MeV/c.  Note that this form of the regulator allows momentum transfers ($\vec p - \vec p'$) to achieve values in the range up to $2  \Lambda_{N3LO}$.  Can one make an  estimate of the uv regulator scale of the Idaho N$^3$LO interaction which is more appropriate to the discrete HO basis of this study?  An emulation of this interaction in a harmonic oscillator basis uses  $\hbar\omega = 30$ MeV and $N_{max} = N=20$  \cite{Barnea2}. Nucleon-nucleon interactions are defined in the relative coordinates of the two-body system so one should calculate $\Lambda^{NN} =\sqrt{m (N + 3/2)\hbar\omega} $ with the $\it reduced$ mass $m$ rather than the nucleon mass $m_N$ appropriate for the single-particle states of the model space.  Taking this factor into account, the successful emulation of the  Idaho N$^3$LO interaction in a HO basis suggests that $\Lambda^{NN}\sim$ 780 MeV/c, consistent with the figure. 

For $\Lambda < \Lambda^{NN}$ there will be missing contributions of size $\vert (\Lambda-\Lambda^{NN})/\Lambda^{NN}\vert$  so ``plateaus" develop as $\lambda\rightarrow 0$  revealing this missing contribution to $\vert\Delta E/E\vert$.  The ``plateau" is not flat as $\lambda\rightarrow 0$ and, indeed, rises significantly with decreasing $\Lambda <\Lambda^{NN}$.  This suggests that corrections are needed to  $\Lambda$ and $\lambda$ which are presently defined only to leading order in $\lambda/\Lambda$.   We hope to learn if higher-order  corrections  can be determined by the data in a future study.

\newpage

\begin{figure}[ht]
\includegraphics[width=16cm,height=12cm,clip]{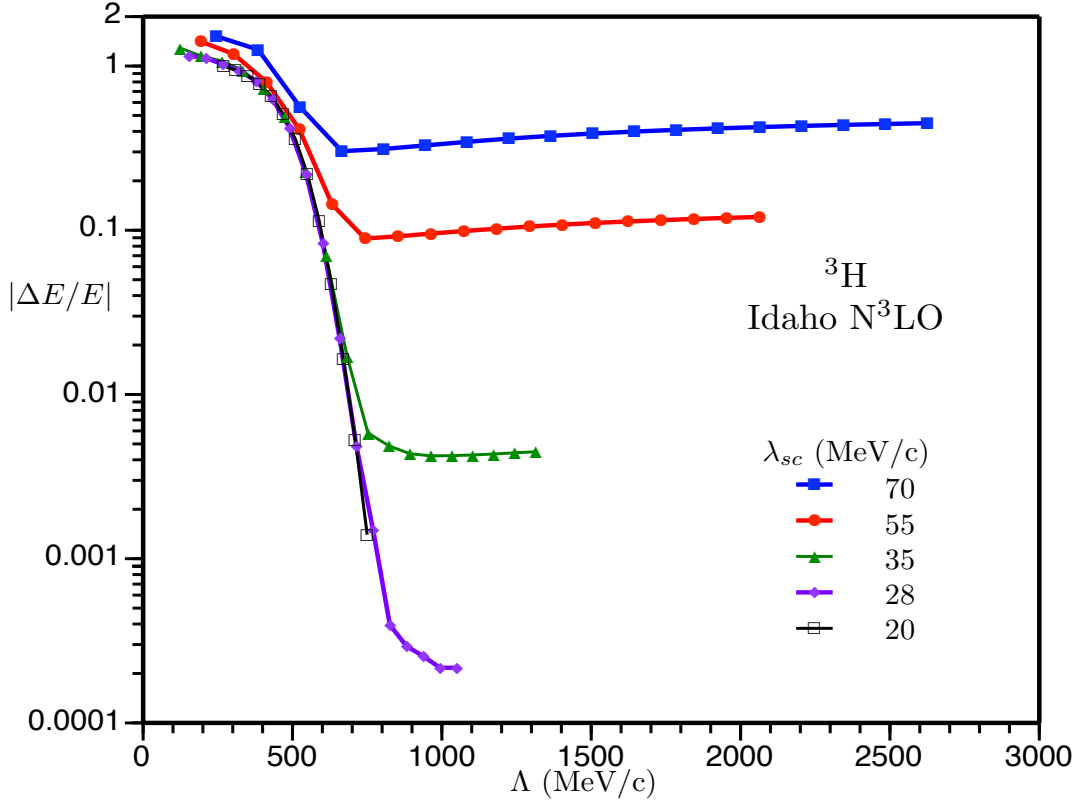}
\caption{(Color online)  Dependence of the ground-state energy of $^3$H (compared to a converged value-see text) upon the uv momentum cutoff $\Lambda$ for different values of  the ir momentum cutoff $\lambda_{sc}$. Curves are not fits but simple point-to-point line segments 
to guide the eye. }
\label{fig:4}
\end{figure}

Now we turn to the second pair of cutoffs  of ($\Lambda,\lambda_{ir}$) and display in Figure 4 the analogue of Figure 2 except that this time  $\lambda_{ir}\equiv \lambda_{sc}  =  \sqrt{m_N \hbar\omega/(N + 3/2)} $.   
For fixed $\lambda_{sc}$, $\vert\Delta E/E\vert$ does not go to zero with increasing $\Lambda$, and indeed even appears to rise for  fixed $\lambda_{sc}\geq 35$ MeV/c and $\Lambda\geq 800$ MeV/c. Such a plateau-like  behavior was attributed in Figure 3  to a uv regulator scale characteristic of the $NN$ interaction.  Can the  behavior of Figure 4 also be explained by  a  ``missing contributions" argument; i.e. an argument based upon $\lambda_{sc}\leq\lambda^{NN}_{sc}$ where  $ \lambda^{NN}_{sc}$ is a second characteristic ir regulator scale implicit in the $NN$ interaction itself?  One can envisage such an ir cutoff as related to the lowest energy configuration that the $NN$ potential could be expected to describe.  For example, the inverse of the $np$ triplet scattering length of 5.42 fm corresponds to a low-energy cutoff of about 36 MeV/c.  Realistic $NN$ potentials such as Idaho N$^3$LO and JISP16 do fit these low-energy scattering parameters  well.  The previously mentioned  emulation of the Idaho N$^3$LO interaction in a harmonic oscillator basis \cite{Barnea2}  has $\lambda^{NN}_{sc}\sim$ 36 MeV/c.  As we shall see later, the fit to low-energy $NN$ data of JISP16 implies a $\lambda^{NN}_{sc}\sim$ 63 MeV/c. The factor of two difference may simply be a reflection of the fitting procedures and appears to be within the range of uncertainty of our argument.  In any event, the behavior of our results in Figure 4 is not inconsistent with this concept of an inherent ir regulator  scale implicit in the $NN$ interaction.

Having introduced a scale  $ \lambda^{NN}_{sc}$,   we continue our discussion of Figure 4.  As fixed $\lambda_{sc}$ requires $\hbar\omega/N$ to be constant and $N\leq36$, small values of fixed $\lambda_{sc}$ are linked with small values of $\Lambda$.  Having said  that, 
 we see that 
  $\vert\Delta E/E\vert\rightarrow 0$ with increasing $\Lambda$ for $\lambda_{sc}=20$ MeV/c  and perhaps $\lambda_{sc}=28$ MeV/c for the values of $\Lambda$ available to the calculation.  At fixed $\lambda_{sc}\geq\lambda^{NN}_{sc}$ and increasing $\Lambda$, once $\Lambda >\Lambda^{NN}$,  a ``plateau"  will develop since no new contributions to  $\vert\Delta E/E\vert$ exist for $\Lambda>\Lambda^{NN}$.  In the figure, the plateau appears to start at $\Lambda\sim 700-900$ MeV/c, consistent with the estimate of $\Lambda_{NN}\sim$ 780 MeV/c for this $NN$ interaction. The plateau in $\vert\Delta E/E\vert$  for larger  fixed $\lambda_{sc}$ is higher than the plateau for small fixed $\lambda_{sc}$ since more contributions to $\vert\Delta E/E\vert$ are missing from the infrared region.   
  Again we observe the plateau rises with increasing $\Lambda$ and 
  this behavior may be a sign that corrections are needed to  $\Lambda$ and $\lambda_{sc}$ which are presently defined only to leading order  in $\lambda_{sc}/\Lambda$. However,  for $\lambda_{sc}\leq\lambda^{NN}_{sc}$  and $\Lambda 
< \Lambda^{NN}$ the results converge to a single curve at the left of  this figure.  It is remarkable that this curve persists to quite low $\Lambda$ values.  This means that $\vert\Delta E/E\vert$ becomes insensitive to $\lambda_{sc}$ for low $\Lambda$ if $\lambda_{sc}$ is low enough.  Later on we will demonstrate that this curve can be quite well described by a Gaussian, a result which persists for other s-shell nuclei.  But we will see in the next figure that one has not yet captured the uv region at these low values of $\Lambda$.

\begin{figure}[ht]
\includegraphics[width=16cm,height=12cm,clip]{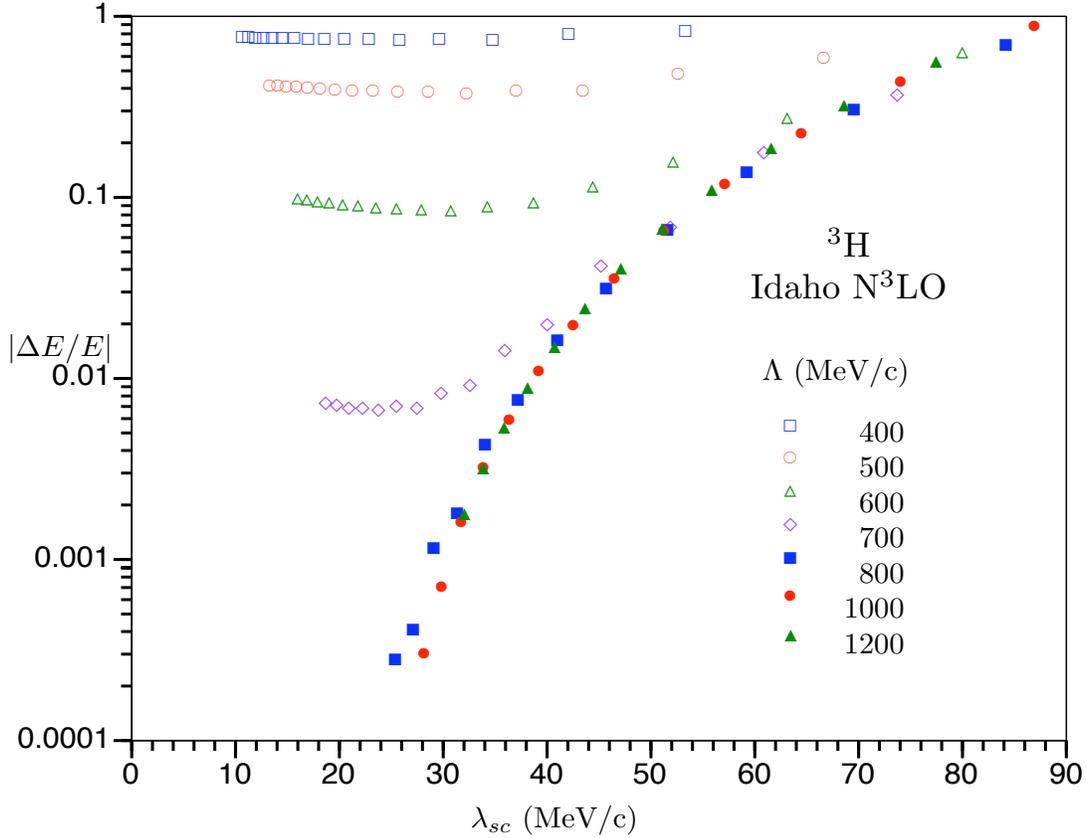}
\caption{(Color online)  Dependence of the ground-state energy of $^3$H (compared to a converged value-see text) upon the ir momentum cutoff $\lambda_{sc}$ for fixed $\Lambda$. }
\label{fig:5}
\end{figure}

Figure 5 is the analogue to Figure 3: only the variable on the x-axis changes from $\lambda$ to $\lambda_{sc}  =  \lambda^2/\Lambda$.  For $\Lambda < \Lambda^{NN}\sim 780$ MeV/c the missing contributions  and resulting ``plateaus" are as evident as in Figure 3. The tendency of these plateaus to rise as $\lambda_{sc}\rightarrow 0$ again suggests
 a refinement is needed to this first-order definition of the cutoffs.   Around  $\Lambda\sim 600$ MeV/c and above the plot of $\vert\Delta E/E\vert$ versus $\lambda_{sc}$ in Figure 5  begins to suggest a universal pattern, especially at large $\lambda_{sc}$.
For $\Lambda\sim 800$ MeV/c and above the pattern defines a universal curve for all values of  $\lambda_{sc}$.  This is  the region  where $\Lambda\geq\Lambda^{NN}$indicating that nearly all of the ultraviolet physics set by the potential has been captured.
Such a universal curve suggests that $\lambda_{sc}$ could be used for extrapolation to the ir limit, provided that $\Lambda$ is kept large enough to capture the uv region of the calculation.  Figure 5 is also the motivation for our appellation $\lambda_{sc}$, which we read as ``lambda scaling", since  this figure exhibits the attractive scaling properties of this regulator.  


\begin{figure}[ht]
\includegraphics[width=16cm,height=12cm,clip]{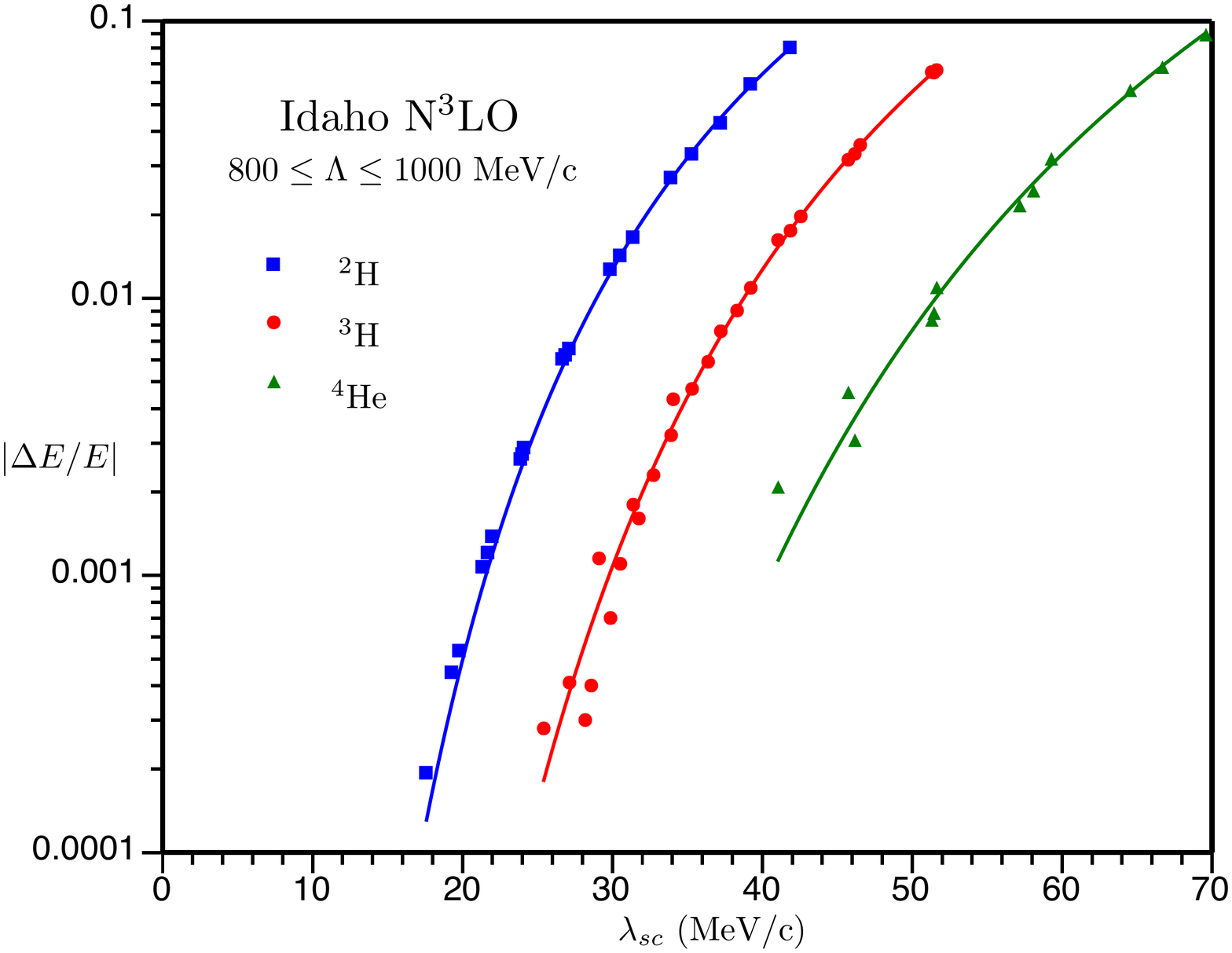}
\caption{(Color online)   Dependence of the ground-state energy of three s-shell nuclei (compared to a converged value-see text) upon the ir momentum cutoff $\lambda_{sc}$ for  $\Lambda$ above the $\Lambda^{NN} \approx 780$ MeV/c set by the $NN$ potential . }
\label{fig 6}
\end{figure}

For Figure 6, we take advantage of the ``saturation" of the uv region by binning all results with $\Lambda\geq 800$ MeV/c.  They do indeed fall on a universal curve for each nucleus shown, indicating that one can use this universal behavior for an extrapolation which is independent of A.
 The curves  are offset  for the three nuclei but otherwise appear similar.  The points can be  fit by the function $y = a\exp(-b/\lambda_{sc})$ with $b\approx 20-40$ MeV/c at A=2,3,4.


\begin{figure}[ht]
\includegraphics[width=16cm,height=12cm,clip]{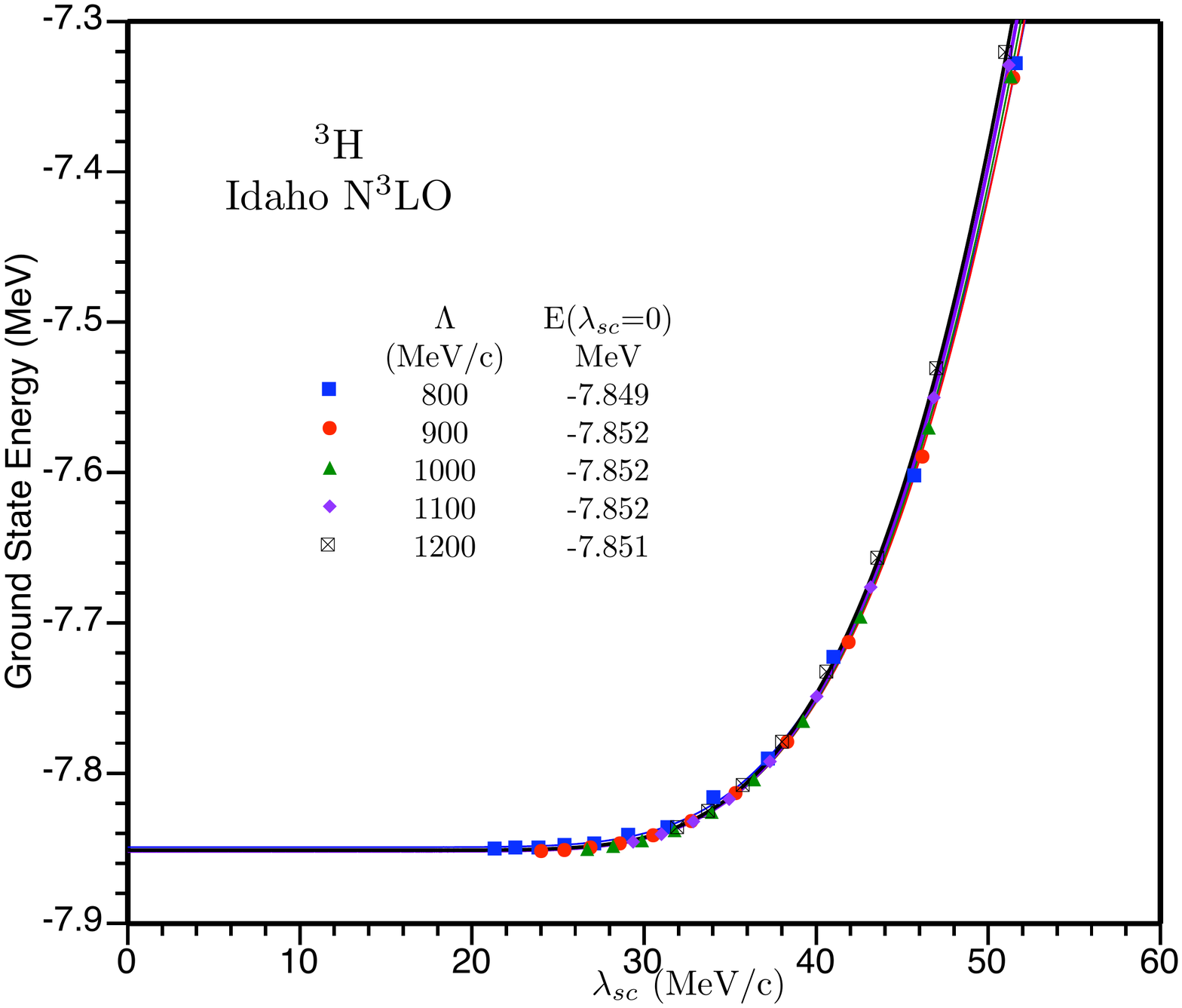}
\caption{(Color online) The ground state energy of $^3$H calculated at five fixed values of $\Lambda$ and variable $\lambda_{sc}$. The curves are fits to the points and the functions fitted are used to extrapolate to the ir limit $\lambda_{sc}= 0$. }
\label{fig:7}
\end{figure}

Finally we utilize the scaling behavior displayed on Figures 5 and 6 to suggest an extrapolation procedure which we demonstrate in Figure 7, again concentrating on $^3$H and the Idaho N$^3$LO potential.    The extrapolation is performed by a fit of an exponential plus a constant to each set of results at fixed  $\Lambda$.  That is, we fit the ground state energy with three adjustable parameters using the relation   \( E_{gs}(\lambda_{sc}) = a \exp(-b/\lambda_{sc}) + E_{gs}(\lambda_{sc}=0)  \). It should be noted that our five extrapolations in Figure 7 employ an exponential function whose argument
$1/\lambda_{sc}  =  \sqrt{(N + 3/2)/(m_N \hbar\omega)}$ is proportional to $\sqrt{N/(\hbar\omega)}$ and is therefore distinct from the popular extrapolation with an exponential in $N_{max}$ ($=N$ for this $s$-shell case) \cite {NavCau04,  Maris09, NQSB, Jurgenson, Bogner,Forssen1, Forssen2}.
 The mean and standard deviation of the five values of $E_{gs}(\lambda_{sc}=0)  $ were 
 $-7.8511$ MeV  and  0.0011 MeV, respectively, as suggested by Figure 7  in which the overlap of the five separate curves cannot be discerned.  For calibration of our scheme, we recall that the accepted value for the ground state of  $^3$H with this potential is $-7.855$ MeV from a 34 channel Faddeev calculation \cite{IdahoN3LO}, $-7.854$ MeV from a hyperspherical harmonics expansion \cite{Kievsky08}, and $-7.85(1)$ from a NCSM calculation \cite{NavCau04}.

This extrapolation procedure with a large and fixed $\Lambda  =  \sqrt{m_N (N + 3/2)\hbar\omega}$ and taking $\lambda_{sc}  =  \sqrt{m_N \hbar\omega/(N + 3/2)}$ toward the smallest value allowed by computational limitations treats both $N$ and $\hbar\omega$ on an equal basis.  
For example, the extrapolation at fixed $\Lambda  =1200$ MeV/c employs values of $\hbar\omega$ from 41 to 65 MeV and $N=22-36$.  The one at fixed $\Lambda  =800$ MeV/c employs values of $\hbar\omega$ from 18 to 44 MeV  and   $N=14-36$. The curves of Figure 7  encompass values of $\lambda_{sc}$ between 20 and 52 MeV/c.   We attempted to
quantify the spread in extrapolated values by  fitting only segments of the curves of this figure.  Recall that the smallest value of $\lambda_{sc}$ requires the largest $N$.  Fits to the   segment from $\lambda_{sc} = 20$ MeV/c to $\lambda_{sc} = 40$ MeV/c (always for the five displayed values of fixed $\Lambda$) resulted in a mean of $-7.8523$ MeV and standard deviation of 0.0008 MeV.  Cutting out the left hand parts of the curves and fitting only  from $\lambda_{sc} = 30$ MeV/c to $\lambda_{sc} = 55$  MeV/c gave a mean of $-7.8498$ MeV and standard deviation of 0.0022 MeV.  For both these trials a rather large $N$ was needed, ranging from 14 to 36 but the extrapolation is quite stable.  In contrast, values of $\lambda_{sc}$ higher than those shown in  Figure 7, namely from $\lambda_{sc} = 50$ MeV/c to $\lambda_{sc} = 85$ MeV/c, require fewer computational resources ($N=8-22$).  The extrapolations have a mean  and standard deviation of $-7.792$ MeV and 0.042 MeV, still not so far away from the accepted value of $-7.85$ MeV.

 
 \begin{figure}[ht]
\includegraphics[width=16cm,height=12cm,clip]{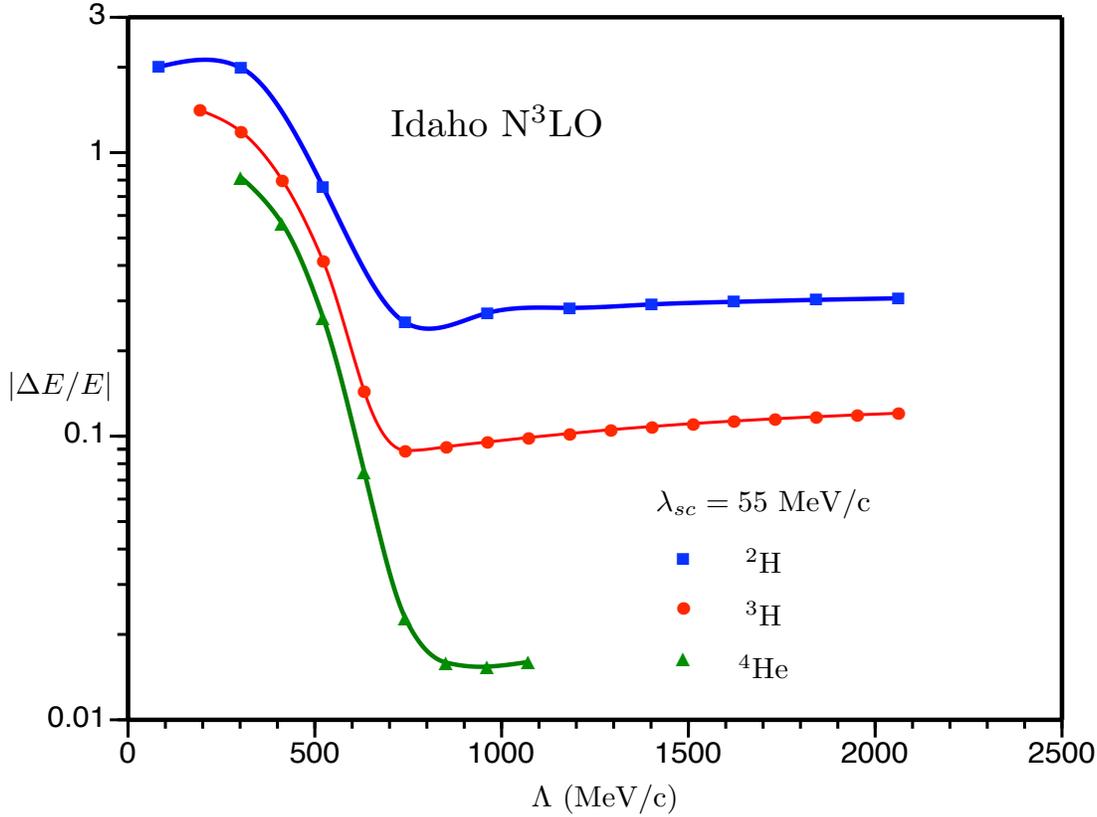}
\caption{(Color online)   Dependence of the ground-state energy of three s-shell nuclei (compared to a converged value-see text) upon the uv  momentum cutoff   $\Lambda$ for $\lambda_{sc}$ above the $\lambda^{NN}_{sc} \approx 36$ MeV/c set by the $NN$ potential.  Curves are not fits but spline interpolations 
to guide the eye. }
\label{fig:8}
\end{figure}

In Figure 8 we return to one of the curves of Figure 4 to examine the dependence of the binding energy of three nuclei upon the uv regulator when the putative ir regulator is held fixed  ($\lambda_{sc} = 55$ MeV/c).   At fixed  $\lambda_{sc}\geq\lambda^{NN}_{sc}\sim 36$ MeV/c,  and increasing $\Lambda$, once $\Lambda >\Lambda^{NN}$,  a ``plateau"  will develop since no new contributions to $\vert\Delta E/E\vert$ exist for $\Lambda>\Lambda^{NN}\sim 780$ MeV/c.     The new feature of this figure is that the ``plateau"  of the nucleus $^2$H is above that of $^3$H (taken from figure 4) which is in turn above that of $^4$He.  This suggests  that $\Lambda_{NN}$ is not the only regulator scale needed to explain the dependencies upon  $\Lambda$ and $\lambda_{sc}$.  Figure 8 introduces another scale - the role of binding momentum (Q) of a nucleus.
The scale $Q$ has been used recently in EFT treatments of pion-deuteron scattering at threshold \cite{Beane}.  The idea is to take the small binding energy of the deuteron explicitly into account as one attempts to develop a consistent power counting for an EFT of pion-nucleus scattering lengths.  The  extension of the definition of Q to more massive nuclei can take alternate forms: $Q = \sqrt{2 m_N(E/A)}$ where $E/A$ is the binding energy per nucleon, or $Q = \sqrt{2 \mu \epsilon}$  where $\mu$ is the reduced mass of a single nucleon with respect to the rest of the nucleons in the nucleus and $\epsilon$ is the binding energy with respect to the first breakup channel \cite{Liebig11}.  Clearly the two definitions coincide for the deuteron and for the light nuclei considered here both definitions give similar estimates.  For definitiveness,
we calculate $Q$ according to the formula  $Q = \sqrt{2 \mu \epsilon}$.	This calculation gives $Q(^2$H) $= 46$ MeV/c, $Q(^3$H)$ = 88$ MeV/c,   $Q(^3$He)$ = 83$ MeV/c ,   $Q(^4$He)$ = 167$ MeV/c,  and $Q(^6$He) $= 39$ MeV/c.  The binding momentum of $^6$He is comparable to that of the deuteron because the first breakup channel into $^4$He$ +2n$ is only about 1 MeV above the ground state.

The fractional error plotted in Figure 8 appears to rise slightly from a minimum at  $\Lambda>\Lambda^{NN}\sim 780$ MeV/c  as the uv cutoff  $\Lambda$ increases for each of the three nuclei.   For example in the  $^2$H calculation the $\lambda_{sc}$ cutoff relative to the deuteron binding momentum is $\lambda_{sc}/Q= 1.2$ and the error is rather high,  rising from a minimum of about 25$\%$.  The triton is more bound so that the ratio  is $\lambda_{sc}/Q= 0.62$ and the minimum error is 9$\%$.  The calculation of the tightly bound $^4$He ($\lambda_{sc}/Q= 0.33$) has the smallest error of less than  2$\%$, but even that error appears to rise as $\Lambda$ increases to the limits of the present calculation.   That is, as  $Q$ increases at fixed $\lambda_{sc}$ (at high enough $\Lambda$) the error due to the $\lambda_{sc}$ cutoff is lower.  It is natural to expect that the many-body dynamics   enters at some level  and sets additional scales beyond the $NN$-interaction scales.

\newpage 
  \begin{figure}[ht]
\includegraphics[width=16cm,height=12cm,clip]{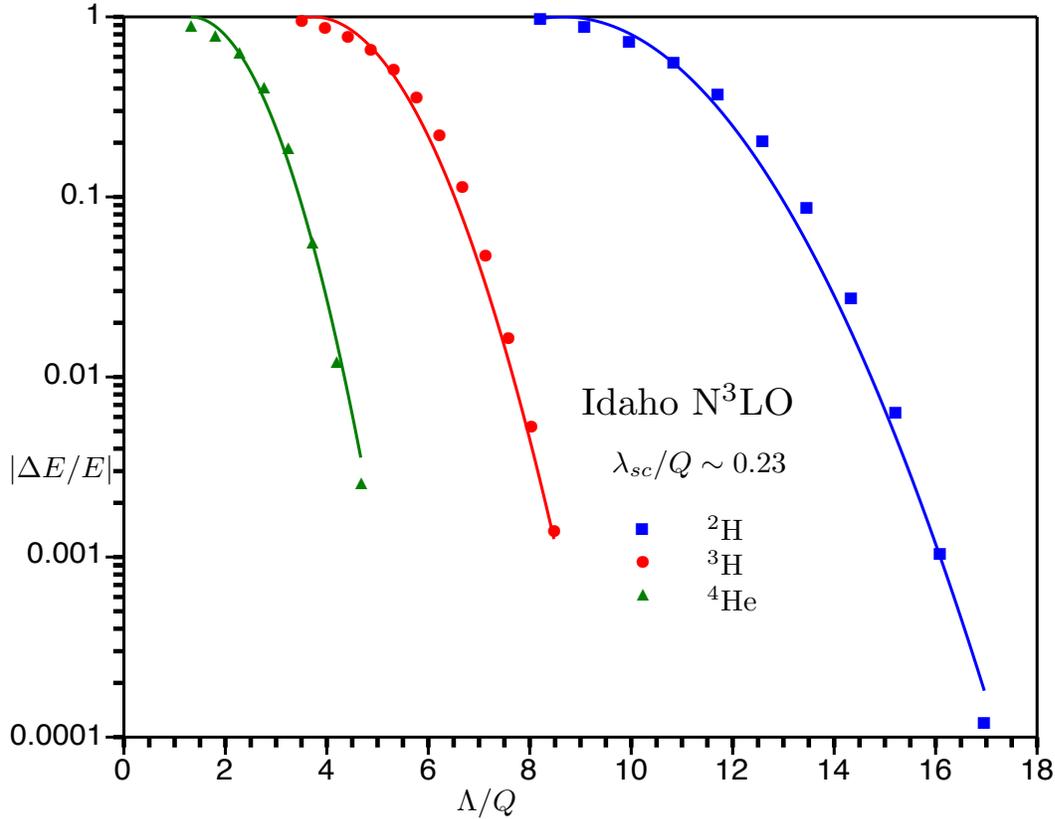}
\caption{(Color online)  Dependence of the ground-state energy of three s-shell nuclei (compared to a converged value-see text) upon the uv  momentum cutoff   $\Lambda$ for $\lambda_{sc}$ below the $\lambda^{NN}_{sc} \approx 36$ MeV/c set by the $NN$ potential. The data are fit to Gaussians. }
\label{fig:8}
\end{figure}

A final comparison of three s-shell nuclei selects the low end of the ir region where $\lambda_{sc}$ is at or below the ir cutoff suggested by the potential.  In Figure 9 all momenta are scaled by the binding momentum $Q$ of the considered nucleus in order to put them on the same plot.  For such low momenta $\lambda_{sc}$, $\vert\Delta E/E\vert$ does go to zero with increasing $\Lambda$ because $\lambda_{sc}\leq\lambda^{NN}_{sc}$, where  $ \lambda^{NN}_{sc}$ is the second  regulator scale of the $NN$ interaction itself.
  For $^3$H, $\lambda_{sc}/Q \sim 0.23$ corresponds to $\lambda_{sc}=20$ MeV/c; the curve can be directly compared with the analogous  curve (black online)  in Figure 4.   For  $^2$H $\lambda_{sc}=10$
 MeV/c  and for $^4$He  $\lambda_{sc}=40$ MeV/c, all values of  $\lambda_{sc}$  are near or below the  second (ir) regulator scale of the Idaho N$3$LO potential  suggested to be  $\sim$ 36 MeV/c.  The largest value of $\Lambda$ plotted is $\sim 861$ MeV/c for  $^2$H, $\sim 746$ MeV/c for $^3$H, and $\sim 780$ MeV/c for $^4$He, all values of $\Lambda$ are near or just above the uv regulator of the Idaho N$3$LO potential suggested to be $\sim 780$ MeV/c.  So this is a plot of {\it low} $\Lambda$ for all the nuclei portrayed.  The  ``high" $\Lambda$  tails of these curves can be fit by  Gaussians (shifted from the origin) in the variable $\Lambda/Q$.  This figure would seem to nicely illustrate the expectations of the theorems of the 1970's \cite{Delves72,Sch72} that the asymptotic rate of convergence does not depend upon the number of particles.

 \begin{figure}[ht]
\includegraphics[width=16cm,height=12cm,clip]{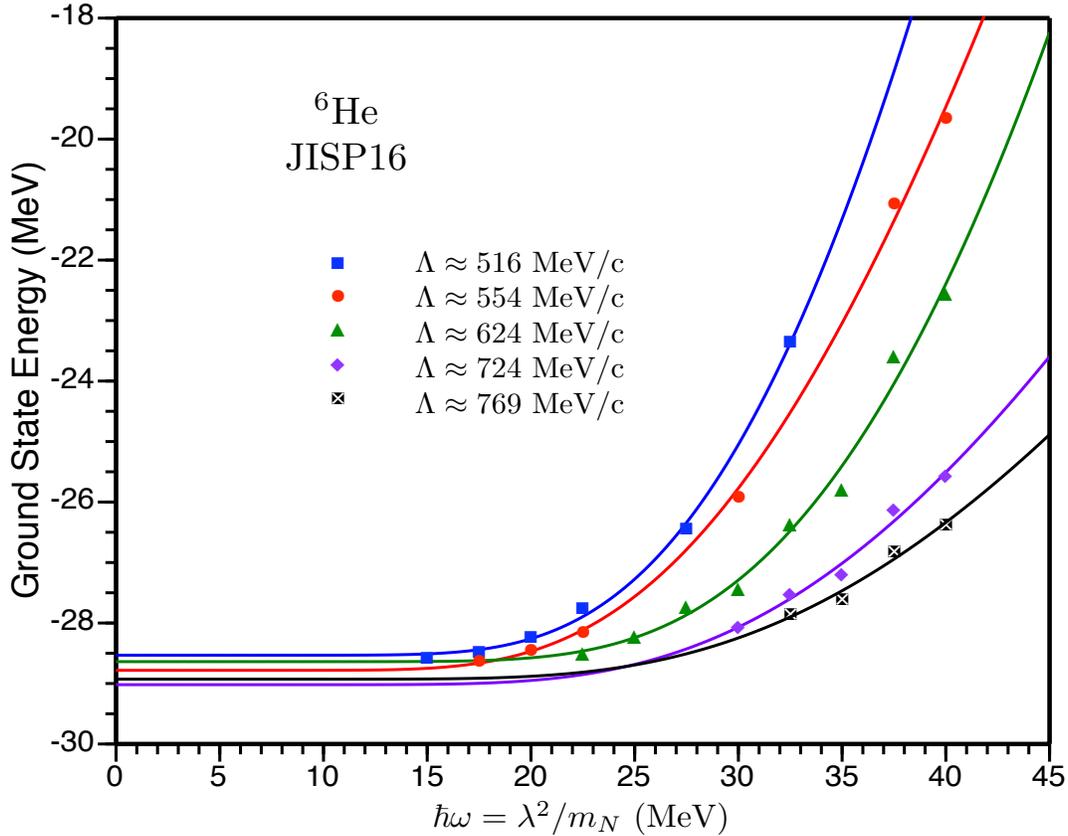}
\caption{ (color online) The ground state energy of $^6$He calculated at five approximate values of $\Lambda$ (see text) and integer values of  $\hbar\omega$. The curves are fits to the points and the functions fitted are used to extrapolate to the ir limit $\lambda= 0$ as in Figure 7.}
\label{fig:9}
\end{figure}

In Figures 4 through 9 we have displayed the features of our results as functions of the
 pair of cutoffs  of ($\Lambda,\lambda_{ir}$) where $\lambda_{ir}\equiv \lambda_{sc}  =  \sqrt{m_N \hbar\omega/(N + 3/2)} $, and demonstrated an extrapolation procedure to the uv and ir limits.   Yet Figure 3 suggests that an extrapolation to the infrared limit could equally well  be made by taking $\lambda \rightarrow 0$  for a fixed large $\Lambda$.   In Figure 10, we demonstrate the features of such an extrapolation by using published results of the JISP16 interaction for the halo nucleus $^6$He \cite{Maris09}.   Its  binding momentum  $Q(^6$He) $= 39$ MeV/c is comparable to that of the loosely  bound deuteron and should provide a severe test of any infrared extrapolation.   

A second reason for considering $^6$He is that it has been studied extensively with the JISP16  $NN$ interaction, both with HO expansion and HH expansion techniques \cite{Yakubovsky}.
 The same set of ground state energy eigenvalues as those plotted in Figure 10 yields an extrapolated value of $ -28.76(9)$ MeV using ``extrapolation A" of Ref.  \cite{Maris09,update}. 
 In ``extrapolation A" one lets  the variable $N\rightarrow\infty$ with a selection procedure for values of $\hbar\omega$  as explained in Ref. \cite{Maris09}. If one refers to the published results \cite{Maris09}   where the largest  $N$ is 15 one finds an ``extrapolation B" method  which lets $N\rightarrow\infty$ at fixed  $\hbar\omega$ to obtain $ -28.69(5)$  MeV in good agreement with the value of $ -28.68(12) $MeV  from``extrapolation A".   A hyperspherical harmonics expansion calculation of $^6$He with the JISP16 potential finds an extrapolated value of $-28.70(13)$ MeV \cite{betadecay}.  This is increased by about 200-300 keV to  $ -28.96(3)$ MeV by a ``hyperspherical harmonics effective interaction"  technique which requires fewer terms in $K$ to reach asymptotic convergence but loses variational character because the induced many-particle  interactions are dropped from the effective interaction \cite {Barnea2}.

  The results of Ref. \cite{Maris09} were obtained with an antisymmetrized many-body wavefunction constructed as   as sum of Slater determinants of single-nucleon wavefunctions depending on single-nucleon coordinates  (and the Lawson method to isolate CM effects) on a mesh of integer ($N,\hbar\omega$). The value of $N$ is, by definition an integer and values of the non-linear variational parameter $\hbar\omega$ were chosen to be increments of 2.5 MeV between 10 and 40 MeV.    To show that  the familiar integer values of $\hbar\omega$ from  Ref. \cite{Maris09} could be directly used in the extrapolation procedures suggested here we mapped  the ground state energy eigenvalues onto the variables ($\Lambda,\hbar\omega = \lambda^2/m_N$) rather than onto the variables ($\Lambda,\sqrt{\hbar\omega} = \lambda/\sqrt{m_N}$).   The largest value of $N$ was 17 ($N_{max}=16$ for this $p$-shell nucleus) \cite{update}.  The extrapolation of Figure 10 is performed by a fit of an exponential plus a constant to the set of results at fixed  $\Lambda$.  The resulting $\Lambda$'s are not then strictly fixed but each point plotted corresponds to a value of $\Lambda$ constant to within $2-5 \%$ of the central value indicated on the graph. 
 The important $S$-wave parts of the JISP16 potential are fit to the data in a space of $N=8$ and $\hbar\omega=40$ MeV.  Therefore this potential has $NN$ regulator scales of $\lambda^{NN}_{sc}\sim$ 63 MeV/c and $\Lambda^{NN}\sim$ 600 MeV/c.  (The value of  $\lambda^{NN}$ associated with this potential is about 200 MeV/c, as can readily estimated from the legend of Figure 2).  But JISP16 seems to be so soft that the ultraviolet region is already captured with $\Lambda\geq 500$ MeV/c, as shown by the top two curves of Figure 10.

  We fit the ground state energy with three adjustable parameters using the relation 
   \( E_{gs}(\hbar\omega) = a \exp(-b/\hbar\omega) + E_{gs}(\hbar\omega=0)  \) five times, once for each ``fixed" value of $\Lambda$.  It is readily seen that one can indeed make an ir extrapolation  by sending $\hbar\omega\rightarrow 0$ with fixed $\Lambda$ as first advocated in Ref.  \cite{EFTNCSM} and that the five ir extrapolations are consistent.  The spread in the five extrapolated values is about 500 keV or about 2$\%$  about the mean of  $-28.78$ MeV.  The standard deviation is 200 keV. 
    

\begin{figure}[ht]
\includegraphics[width=16cm,height=12cm,clip]{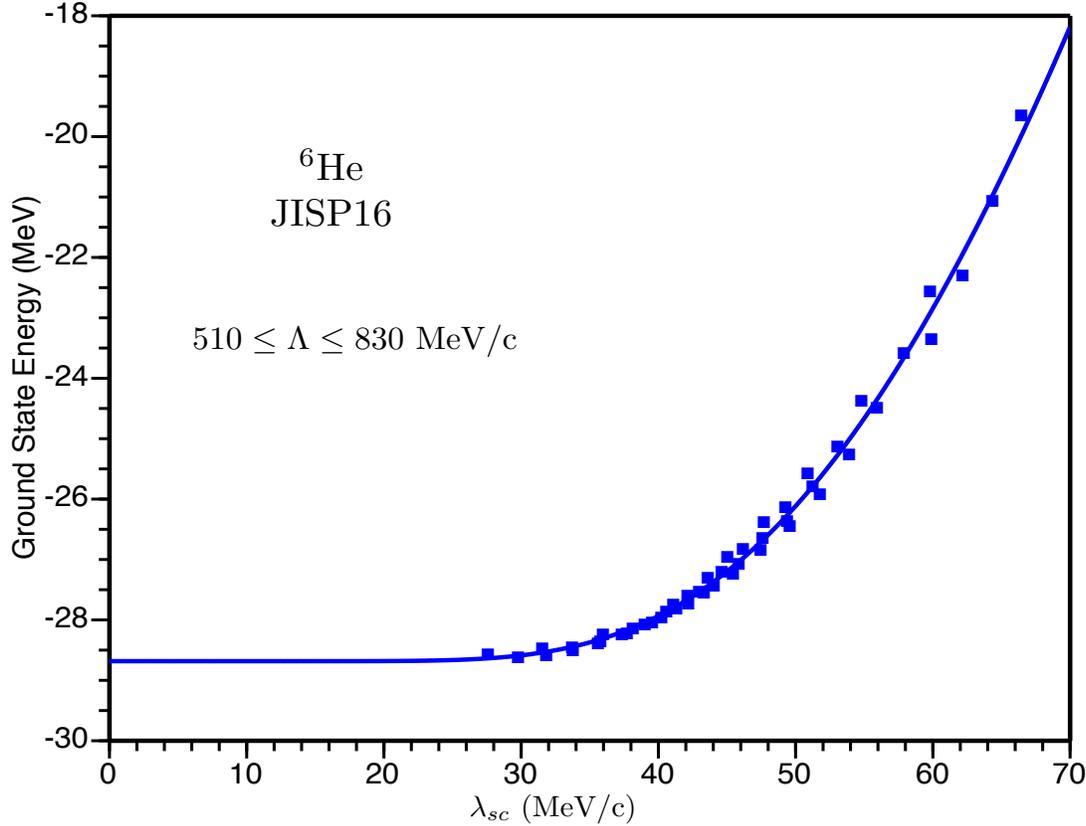}
\caption{(Color online)  The ground state energy of  $^6$He calculated at all values  of $\Lambda\geq 510$ MeV/c and variable $\lambda_{sc}$. The curve is a fit to the points and the function fitted is  used to extrapolate to the ir limit $\lambda_{sc}= 0$}
\label{fig:10}
\end{figure}

    A second (single) extrapolation of the  $^6$He data with $\lambda_{sc} \rightarrow 0$  which uses all calculated energies where   $\Lambda\geq 510$ MeV/c is shown in Figure 11.  As in Figure 7,   we fit the ground state energy with three adjustable parameters using the relation   \( E_{gs}(\lambda_{sc}) = a \exp(-b/\lambda_{sc}) + E_{gs}(\lambda_{sc}=0)  \).  The extrapolated value is $-28.68$ MeV, which agrees well with Figure 10 and the other extrapolated results.   The extrapolation prescription used in Figure 11 employs  values of $\hbar\omega$ from 15 to 40 MeV and a range of $N$ from 7-17.  That is,  all of the information (at $\Lambda\geq 510$ MeV/c) available from these calculations is used in the  $\lambda_{sc}$ extrapolation.  How can one estimate an uncertainty from such a single extrapolation?  Looking at the scatter of the points about the fitted curve is instructive but not quantitative.  If we bin the 48 points of Figure 11 into the same bins of ``constant" $\Lambda$ as in Figure 10, we find (not shown) five extrapolations with a mean of $-28.58$ MeV and standard deviation of 0.06 MeV.  Another possible way of breaking up this single extrapolation is more in the spirit of the earlier extrapolations of Delves and successors.  If we map from ($N,\hbar\omega$) onto ($\Lambda,\lambda_{sc}$)  holding $N$ fixed we get another set of extrapolations, those for $N=7,9,11,13,15,17$ with concomitant smallest $\lambda_{sc} = 60,50,41, 36,32,28$ MeV/c.  Of these six extrapolations, only those with $N\geq 13$ are consistent with the extrapolation which uses the full 41 points.  This is to be expected, as one needs a large $N$ before the convergence ``starts to behave".  The mean of the three extrapolations with $N\geq 13$ is   $-28.54$ MeV and standard deviation is 0.11 MeV.  Concentrating only on large $N$ in this naive manner gives a worse extrapolation compared to accepted extrapolated ground state energies.  It would appear that it is advantageous to take advantage of the scaling properties of $\lambda_{sc}$ for all values of  the uv regulator  large enough to capture the uv limit.  In that case, as seen in Figure 11,  even results with low $N$ (and therefore large $\lambda_{sc}$) can usefully stabilize and bound an extrapolation to the ir limit.  A rough estimate of the uncertainties of this extrapolation of figure 11  would then be $ -28.68(22)$  MeV. 

In conclusion, our extrapolations in the ir cutoff $\lambda$ of $-28.78(50)$ MeV or  the ir cutoff $\lambda_{sc}$ of $28.68(22)$ MeV are consistent with each other and with the independent calculations.

\section{Summary and outlook}
\label{sec:5}
We reviewed  the functional analysis theorems which describe variational calculations of many-body systems made with a trial function expanded in a complete set of known functions.  According to these theorems  the convergence properties of  such a calculation are 
determined by the interaction and by the dimensionless number $\mathcal{N}$ which determines  the truncation at a finite number of basis functions.   Among basis sets, harmonic oscillator (HO) functions are distinguished by ease of separation of relative and center of mass coordinates and by the dimensional parameter  $\hbar\omega$  which sets an intrinsic scale.  Motivated by effective field theory studies, one can define  quantities from ($\mathcal{N},\hbar\omega$)  forming  ultraviolet (uv) and infrared(ir) momenta  that act as cutoffs that characterize the model space just as does ($\mathcal{N},\hbar\omega$).    Extending both the uv cutoff to infinity and the ir cutoff to zero  is prescribed for a converged calculation. There have been two alternate definitions of the ir cutoff;  $\lambda = \sqrt{m_N \hbar\omega}$ and $\lambda_{sc}  =  \sqrt{m_N \hbar\omega/(N + 3/2)}$.  Note that $\lambda_{sc}  =  \lambda^2/\Lambda$  where $ \Lambda  =  \sqrt{m_N (N + 3/2)\hbar\omega} $ is the uv cutoff as usually defined.  We calculated the ground state energy of light nuclei with the ``bare" and ``soft" $NN$ interactions Idaho N$^3$LO and JISP16.  We investigated the behaviors of the uv and ir regulators of model spaces used to describe  $^2$H, $^3$H, $^4$He and $^6$He. 

We obtained fully converged eigenvalues for $^2$H and  $^3$H which were in agreement with other calculations including  those (e.g. Faddeev approach) obtained from  a direct finite difference solution of partial differential equations in many dimensions.  These results  could be used to examine the cutoff dependences of the model spaces ($\Lambda,\lambda$) or ($ \Lambda,\lambda_{sc}$) as one cutoff was held fixed and the other approached its limit.   The examination was based upon the ratio  $\vert\Delta E/E\vert $, defined as $\vert (E(\Lambda,\lambda_{ir}) - E)/E\vert$ where $E$ is the fully converged ground state energy. Both pairs of cutoffs acted as expected when $\Lambda$ was held fixed and $\lambda_{ir}$ tended toward zero; $\vert\Delta E/E\vert \rightarrow 0$ in  Figures 3 and 5 provided $\Lambda$ exceeds a threshold set by the potential.  On the other hand, in both figures drawing an imaginary vertical line at a fixed $\lambda_{ir}$ which crosses the curves shows that the calculation gets better as $\Lambda$ increases.  It is in Figures 2 and 4 that the difference between the two versions of the $\lambda_{ir}$ cutoff become evident.  For the pair 
($\Lambda,\lambda$) in Figure 2 $\vert\Delta E/E\vert \rightarrow 0$ as the uv cutoff increases for all values of $ \lambda$ investigated.   But in  Figure 4  $\vert\Delta E/E\vert $ actually rises as 
$\Lambda\rightarrow\infty$  if $\lambda_{sc}$ is larger than another threshold value evidently set by the potential.  As both cutoffs should be sent to their respective limits for a converged calculation, this behavior does not invalidate the identification of $\lambda_{sc}$ with $\lambda_{ir}$, but it does seem a little peculiar.  Perhaps this behavior signals a need for higher order terms  in $\lambda/\Lambda$  in the definition of $\lambda_{sc}$

In any event, we have introduced a practical extrapolation procedure with $\Lambda\rightarrow\infty$ and $\lambda_{ir}\rightarrow 0$ which can be used when the size of the basis exceeds the capacity of the computer resources as it does for $^4$He and $^6$He and certainly will for any more massive nuclei. We have established that  $\Lambda$
does not need to be extrapolated to $\infty$ but if   $\Lambda>\Lambda^{NN}$ set by the potential one can make the second  extrapolation to zero with either ir cutoff  $\lambda_{sc}$ (see Figures 7 and 11) or $\lambda$ (see Figure 10).  The choice of the scaling cutoff $\lambda_{sc}$ is especially attractive as $\Lambda$ need not be held constant but $\it any$ $\Lambda$ large enough can be used in the ir extrapolation.  Unlike other extrapolation procedures the ones advocated in this paper treat the variational parameters $\mathcal{N}$ and $\hbar\omega$ on an equal footing to extract the information available from sequences of calculations with model spaces described  by ($\mathcal{N},\hbar\omega$)

For the future, we can envisage extending this extrapolation technique to calculating  other properties of nuclei, properties which may or may not be as amenable as are energy eigenvalues to the uv and ir regulators.   The rms point matter radii and the Gamow-Teller matrix element (relevant to $\beta$ decay) of light nuclei are important quantities to calculate reliably for these (and more massive nuclei) \cite{betadecay,H3beta}.  In the nuclear structure folklore, $r^2$ and $D_z$ (the $z$ component of the electric-dipole operator) are of long range and the full GT matrix element, including meson-exchange currents, is of medium range.  The electric dipole polarizabilities of light nuclei are necessary in order to obtain accurate nuclear-polarization corrections for precisely measured transitions involving $S$-waves in one-and two-electron atoms.  The defining relation for the polarizability 
can be converted into a procedure which needs only bound-state quantities and involves the long-range dipole operator $D$  \cite{Stetcu1}.  A convergence analysis of a HO expansion, which lets $N\rightarrow\infty$ at fixed  $\hbar\omega$, for electric dipole polarizabilities of $^3$H, $^3$He and $^4$He obtained faster convergence for lower $\hbar\omega$ than for  the binding energy itself \cite{Stetcu2}.  It would be interesting to learn how the procedure advocated here would work for these problems.  These latter problems, often require not only converged ground state energies, but energies which agree with experiment.  For that, $NNN$ interactions are considered necessary \cite{NO}, thereby leading to a need for more studies of the convergence and extrapolation concepts of this paper.

\section{Acknowledgements}
This study was conceived  and initiated at the National Institute for Nuclear Theory's program 
{\it Effective Field Theories and the Many-Body Problem} in the spring of 2009. We have benefitted greatly from discussions with Sean Fleming, Bruce Barrett, Sigurd Kohler, Jimmy Rotureau,  Chieh-Jen Yang and Eric Jurgenson.   SAC thanks Wayne Polyzou for useful discussions.  Michael Kruse gratefully  acknowledges hospitality at the Nuclear Theory and Modeling Group of Lawrence Livermore National Laboratory (LLNL) during part of this effort.  We are grateful to  Petr Navr\'atil for generously allowing us to use his No-Core Shell Model Slater Determinant Code and his $\it manyeff$ code for many of the calculations.

This work was supported in part by USDOE Division of Nuclear Physics
grants DE-FG-02-87ER40371 (Intermediate/High Energy Nuclear Physics),   DE-FC02-09ER41582 (Building a Universal Nuclear Energy Density Functional (UNEDF) SciDAC Cooperative Agreement) to  ISU and DE-FG02-04ER41338 (Effective Theories of the Strong Interaction) to Arizona.  Additional support came from NSF awards 0904782 (Collaborative Research: Taming the scale explosion in nuclear structure calculations) to ISU and 0854912 (New Directions in Nuclear Structure Theory) to Arizona. 

The calculations with the Idaho N$^3$LO $NN$ interaction  were done with
 allocations of computer time from the UA Research Computing High Performance Computing (HPC) and High Throughput Computing (HTC) at the University of Arizona and from the LLNL
institutional Computing Grand Challenge program.  Numerical calculations
have been performed in part at the LLNL LC facilities supported
by LLNL under Contract No. DE-AC52-07NA27344.  Computational resources for calculations with the JISP16 $NN$ interaction were provided by the National Energy Research Supercomputer Center (NERSC) supported by the Office of Science of USDOE.  The analysis of this paper was made with a Mini-Mac supported by DE-FG02-04ER41338.




\begin{thebibliography}{100}
\bibitem{update}  We used the original  values of $\hbar\omega$ and  extended the data base in $(N,\hbar\omega)$ to $N=17$, one step of $N$ higher than the published results in Ref. \cite{Maris09}.
\end{thebibliography}
\end{document}